\documentclass[preprint1]{aastex6}
\usepackage{amsmath}
\usepackage{calligra}
\usepackage{lineno}
\slugcomment{}
\def\lta{\,\raise 0.3 ex\hbox{$ < $}\kern -0.75 em
 \lower 0.7 ex\hbox{$\sim$}\,}
\def\gta{\,\raise 0.3 ex\hbox{$ > $}\kern -0.75 em
 \lower 0.7 ex\hbox{$\sim$}\,} 

\newcommand{\be}{\begin{equation}}
\newcommand{\ee}{\end{equation}}

\shorttitle{Distributions of Short-Lived Radionuclides} 
\shortauthors{Fatuzzo \& Adams}

\begin{document} 

\title{Theoretical Distributions of Short-Lived Radionuclides \\
for Star Formation in Molecular Clouds} 

\author{Marco Fatuzzo}
\affil{Department of Physics, Xavier University, Cincinnati, OH 45207; \\
fatuzzo@xavier.edu.}
\author{Fred C. Adams}
\affil{Department of Physics, University of Michigan, MI, 48109  \\
Department of Astronomy, University of Michigan, MI, 48109  \\
fca@umich.edu.}

\begin{abstract}

Short-lived radioactive nulcei (half-life $\tau_{1/2}\sim1$ Myr)
influence the formation of stars and planetary systems by providing
sources of heating and ionization. Whereas many previous studies have
focused on the possible nuclear enrichment of our own Solar System,
the goal of this paper is to estimate the distributions of
short-lived radionuclides (SLRs) for the entire population of stars
forming within a molecular cloud. Here we focus on the nuclear species
$^{60}$Fe and $^{26}$Al, which have the largest impact due to their
relatively high abundances. We construct molecular cloud models and
include nuclear contributions from both supernovae and stellar
winds. The resulting distributions of SLRs are time dependent with
widths of $\sim3$ orders of magnitude and mass fractions
$\rho_{\scriptstyle SLR}/\rho_\ast\sim10^{-11}-10^{-8}$. Over the
range of scenarios explored herein, the SLR distributions show only
modest variations with the choice of cloud structure (fractal
dimension), star formation history, and cluster distribution. The most
important variation arises from the diffusion length scale for the
transport of SLRs within the cloud. The expected SLR distributions are
wide enough to include values inferred for the abundances in our Solar
System, although most of the stars are predicted to have smaller
enrichment levels. In addition, the ratio of $^{60}$Fe/$^{26}$Al is
predicted to be greater than unity, on average, in contrast to Solar
System results.  One explanation for this finding is the presence of
an additional source for the $^{26}$Al isotope. 

\end{abstract} 


\section{Introduction} 
\label{sec:intro} 

Star formation environments can influence their constituent stars and
planetary systems through a variety of channels, including dynamical
interactions and background radiation fields, as well as exposure to
high energy particles and radioactive nuclei (see, e.g.,
\citealt{adams2010,pfalzner2015,parker2020}). Most stars form within
embedded clusters \citep{ladalada,porras}, and the immediate cluster
properties largely determine the degree of dynamical disruption
\citep{adams2006,malmberg2007,zwart2009,pfalzner2013} and the
intensity of the background radiation
\citep{fatadams2008,leehop,parker2021}.  The particle contribution
includes both high energy cosmic radiation and short-lived radioactive
nuclei, which are produced via supernovae and stellar winds
\citep{vasileiadis2013,adams2014,lichtenberg16,kuffmeier2016,
  nicholson2017,kaur2019}. The accelerated particles can propagate
beyond the immediate cluster environment and thereby influence
additional star forming regions within the molecular cloud. This paper
focuses on the large-scale distribution of the short-lived
radionuclides (SLRs), which affect star and planet formation through
several mechanisms (as outlined below). The goal is to determine the
distribution of enrichment levels for SLRs over the entire population
of stars and planetary systems forming within the molecular cloud.

Short-lived radioactive nuclei can provide a significant source of
energy for the process of planet formation \citep{urey1955}. As one
example, the additional energy from radioactive decay leads to the
melting of sufficiently large planetesimals. The resulting molten
bodies are more easily differentiated than their solid counterparts
and their volatile components (including water) are more readily
removed (e.g., \citealt{lichtenberg19,reiter2020}). The decay products
from radioactive nuclei represent a substantial ionization source.
The degree of ionization affects the coupling of disk material to
magnetic fields, the onset of magneto-rotational-instability (MRI;
\citealt{balbus1991}), the resulting viscosity of the disk, and
ultimately disk accretion. Ionization also affects the chemical
constituents of the disk \citep{cleeves2013,cleevesradio}, and finally
the composition of forming planets.

The abundances of SLRs in our own solar system, as estimated from
meteoritic analysis \citep{meyer2000,dauphas}, are inferred to be
elevated by an order of magnitude relative to values expected for the
background galaxy (e.g., \citealt{begemann,diehl2006,diehl2013}).  For
example, the ratio $^{26}$Al/$^{27}$Al inferred for the interstellar
medium is $8.4\times10^{-6}$ \citep{diehl2006}, compared with values
$\sim5\times10^{-5}$ measured for the early Solar System
\citep{macpherson1995}.  The explanation for the source of these SLRs
poses an interesting problem regarding the formation of our solar
system \citep{cameron1977}, and the necessity of SLR enrichment places
strong constraints on its birth environment \citep{adams2010}.
Several different sources for the SLRs of our solar system have been
proposed (see the review of \citealt{lugaro2018}), including a local
supernova in the solar birth cluster (e.g.,
\citealt{clayton,adams2001,ouellette,pan2012} and many others),
spallation in the interstellar medium \citep{desch2010}, local
protostellar spallation \citep{shu1997,lee1998}, and distributed
enrichment from many different supernova within the molecular cloud
that formed the Sun \citep{gounelle2012,gounelle2015}.

Much of the aformentioned work has primarily focused on finding an
explanation for the degree of radioactive enrichment of our Solar
System. Because many different nuclei are observed via meteoritc
analysis (including $^{10}$Be, $^{26}$Al, $^{36}$Cl, $^{41}$Ca,
$^{60}$Fe, $^{53}$Mn, $^{107}$Pd, and $^{129}$I), the required
ensemble of abundance SLR ratios places tight constraints on
enrichment scenarios. For completeness, we note that the same
supernovae that provide nuclear enrichment could also trigger the star
formation event that led to the solar system \citep{boss2017}. 

On the other hand, it remains possible for the Solar System SLR
abundances to be closer to those expected for star formation in clouds
(see the discussion of \citealt{jura2013,young2014,young2016,young2020}, 
and references therein). If the galactic supply of $^{26}$Al is
distributed only within molecular clouds, rather than throughout all
of the gas in the interstellar medium, then the galaxy-wide abundance
ratio becomes $^{26}$Al/$^{27}$Al $\sim2-3\times10^{-5}$, roughly
comparable to Solar System values. In addition, white dwarf stars that
are polluted by rocky bodies show evidence that the impactors are
differentiated, which implies heating at levels comparable to those
expected from (inferred) Solar System abundances of $^{26}$Al. As a
result, the observed Solar System SLRs may not be anomalous.

Independent of the SLR inventory of our Solar System, this paper takes
a wider view and seeks to estimate the distribution of radioactive
enrichment levels for the entire population of forming stars and
planets within a molecular cloud. For purposes of heating and
ionization, the particular nuclei $^{60}$Fe and $^{26}$Al provide the
dominant contribution and are the focus of this work (although we also
briefly discuss $^{36}$Cl and $^{41}$Ca).  Both of these principal
SLRs are synthesized in supernovae, as considered here. In addition,
stellar winds from evolved stars provide an additional source of
$^{26}$Al, which is also included. We ignore spallation\footnote{Note
that spallation can produce $^{26}$Al but not $^{60}$Fe.} and other
local sources (although their contributions can be considered
separately).

This work is the analog of a previous paper \citep{adams2014}, where
we estimated the distributions of radioactive abundances for the
scenario where circumstellar disks are enriched by supernovae
exploding within their birth clusters (see related work by
\citealt{lichtenberg16}, \citealt{nicholson2017}). In this paper, we
estimate the corresponding distributions for the radioactive
enrichment scenario where the SLRs propagate through the entire star
forming region.  Instead of focusing on specific cases that account
for the full range of isotopes inferred for our solar system (e.g.,
\citealt{gounelle2012},\citealt{gounelle2015}), the goal of this work
is to construct the distributions of $^{26}$Al and $^{60}$Fe for all
stars forming in the cloud. These distributions can then be used to
assess the degree of radioactive heating, ionization, and other
processes that are present during planet formation across the entire
stellar population.  Previous work along these lines has been carried
out using detailed numerical simulations of molecular clouds
\citep{vasileiadis2013,kuffmeier2016}. These numerical studies provide
distributions of the SLRs as a function of time for the specific
scenario under consideration. This present work --- which is
complementary to these earlier studies --- considers a parametric
model, which allows for the exploration of a wider range of input
parameters, albeit within the restricted geometries and 
star formation prescriptions of the model. 
We thus determine how the distributions of SLR abundances
vary with the time and spatial dependence of star formation, the
fractal dimension of the molecular cloud, cluster properties,
propagation efficiencies, and other inputs.  Although this paper
focuses on nuclear enrichment on scales of molecular clouds, we note
that additional studies have also considered enrichment on galactic
scales (e.g., \citealt{fujimoto2018,cote2019,kaur2019}). For example,
this latter study estimates the time evolution of the galactic-scale
abundances of SLRs and also constructs a working scenario to explain
the observed isotopic abundances in our Solar System.

Although the goal of this paper is to determine the expected
distributions of SLRs for the general population of stars forming in
molecular clouds, the observed abundances inferred for our Solar
System provide a useful reference point.  For the isotope $^{26}$Al,
meteoretic studies imply a mass fraction $X_{26}\sim4\times10^{-9}$
(starting with \citealt{cameron1977}).  For the case of $^{60}$Fe,
however, different studies find a wide range of abundances. The
largest estimates are $X_{60}\sim10^{-9}$ with many measurements
indicating much smaller values (compare \citealt{tachibana2006,
moynier2011,telus2012,mishra2014,trappitsch2018}). In this paper, 
when we compare with Solar System values for the SLRs, we use the 
aforementioned $^{26}$Al mass fraction and the high end of the 
inferred $^{60}$Fe mass fraction ($10^{-9}$) for reference values.
One should keep in mind that the latter mass fraction is highly 
uncertain.

This paper is organized as follows. Section \ref{sec:baseline}
formulates our model for the nuclear enrichment of molecular
clouds. We must specify the cloud structure, distribution of star
clusters, star formation history, stellar initial mass function,
nuclear yields, and the mechanism for the propagation of radioactive
material through the cloud. The results for our benchmark model are
then presented in terms of the distributions of enrichment levels
(mass fractions of the SLRs) as a function of time.  Section
\ref{sec:variations} explores how the resulting distributions vary
with the chosen input parameters, including cloud structure, star
formation history, and the propagation of radioactive nuclei.
Although the distributions are wide, with nuclear abundances ranging
over several orders of magnitude, the distributions themselves are
relatively robust. The paper concludes in Section \ref{sec:conclude}
with a summary of our results and a discussion of their implications.

\section{Baseline Model}
\label{sec:baseline} 

\subsection{Stellar Distribution in the Molecular Cloud Environment}

Giant molecular clouds (GMCs) are complex and varied structures (e.g.,
\citealt{dobbs2013}) with evolutionary histories that remain under
investigation. Given the current uncertainties concerning the physical
characteristics of these environments, we first develop a baseline
model that uses standard assumptions about the GMC stellar population
and the SLR transport mechanisms. We then consider several variations
to our baseline model that explore other viable scenarios in order to
gauge the sensitivity of our results to the physical input parameters.

In all cases, we embed a GMC within a well-define cuboid structure
with a 50 pc body diagonal (e.g., \citealt{heyer2015}).  The cloud is
assumed to contain $10^6\; M_\odot$ of gas out of which $N_{mc}^*=
50,000$ stars form.  The three physical dimensions of the cuboid are
generated by randomly drawing numbers $w_i$ between $e^{0.5}$ and
$e^2$ (where $i =\{x,y,z\}$), and assigning $\ell_i=\ln w_i$ as
relative lengths which are then normalized in accordance to the body
diagonal.  The resulting GMC volumes for this scheme, which range from
6,550 pc${^3}$ to 24,100 pc${^3}$, have an expectation value $\langle
V_{GMC}\rangle\approx 20,000\;{\rm pc^3}$.  We note that a sphere with
the same volume would have a radius of $r \approx 17$ pc.

In our baseline model, the formation of stars occurs within clusters
whose stellar memberships $N$ span from $10^2$ to $10^4$ in accordance
to the distribution function (e.g., \citealt{ladalada}),  
\be
P_c(N)\propto N^{-2}\,.
\label{clusterdist} 
\ee
Clusters are drawn from this distribution function until their
collective stellar content reaches $N^*_{mc}$, with the last drawn
cluster's membership reduced as needed to achieve that outcome.  Each
cluster is assigned a radius through the empirical formula 
\be
R_c = 1\,{\rm pc}\;\left({N\over 300}\right)^{1/3}\;,
\ee
\citep{adams2010}. For our set assumptions, the mean cluster membership
is $\langle N\rangle = 465$, so that a giant molecular cloud is
expected to have $\sim 10^2$ clusters with radii $R_c = 0.7 - 3.2$ pc.
We note, however, that the cumulative probability for finding stars in
clusters with membership size $N$ or smaller is given by the
expression 
\be
P_*(N) = {\ln[N/N_{min}]\over \ln[N_{max}/N_{min}]}\,,
\ee
indicating that stars populate clusters in a logarithmic sense.  In
other words, stars can be found with roughly equal probability in each
decade of cluster membership size $N$.  For our assumed cluster
distribution, half of the GMC stars are expected to belong to clusters
with membership $N\le10^3$.

Following the fractal generating scheme of \cite{goodwin2004}, 
clusters are randomly placed within the cuboid GMC region with
spacial fractal dimension $d$ (with $d = 2$ assumed for our baseline
model).  The main GMC structure is evenly divided into 8
sub-structures, the centers of which are seeded with a ``child" that
has a probability $P = 2^{d-3}$ of maturing.  If a child does not
mature, the sub-structure it was born in becomes inactive, and no
clusters are placed within it.  If the child matures, the
sub-structure is further divided evenly into 8 sub-structures, and the
process is repeated until enough mature children exist to populate the
required number of clusters.  One complication in the adopted scheme
is that the volume of the largest clusters exceeds that of the
sub-structures associated with the final generation of mature
children.  As such, we stray from the process outlined in 
\cite{goodwin2004} by keeping all mature children (as opposed to
keeping only the last generation).  Larger clusters are placed
randomly first at locations with mature children born into
sub-structures that are minimally larger than the cluster volume, and
all higher generation ``active" sub-structures within are removed from
the pool of possible cluster sites.  The process is repeated until all
clusters are placed with no overlapping volume.

In our baseline model, star formation within a given cluster begins at
a time chosen randomly between $t = 0 - 10$ Myrs, and proceeds with
equal probability over a time interval of $\Delta t =2\,{\rm Myrs}$.
The corresponding stellar-birth distribution function for the entire
GMC is therefore given by the piecewise function 
\be 
b(t)=
\left\{ \begin{array}{ll} t/(20\; \text{Myr}^2) &\qquad 0\leq t\leq
  2\;\text{Myr} \\ 1/(10 \;\text{Myr}) &\qquad 2\;\text{Myr}\leq t\leq
  10\;\text{Myr} \\ (12\;\text{Myr}-t)/(20\; \text{Myr}^2) &\qquad
  10\;\text{Myr} \leq t\leq 12\;\text{Myr} \\ 0 &\qquad t >
  12\;\text{Myr}\,.
              \end{array} \right.
\ee
The 12 Myr duration of the star formation epoch is comparable to the
crossing time of the molecular cloud (for transport speed $\sim1$ km/s
and size scale $\sim10$ pc), where this time is characterstic of 
molecular cloud operations (e.g., \citealt{elmegreen2000,dobbs2013}). 

Since SLRs are produced exclusively by massive stars 
($8M_\odot\le M_\ast\le 120 M_\odot$), we adopt an initial mass function 
\be
f(m) = 0.173 \;m^{-2.5}\;,
\ee
(where $m = M_\ast/M_\odot$) that mimics the high-mass portion of the
observed IMF (e.g., \citealt{scalo1998,kroupa2001}). The normalization
is chosen so that the distribution has a given probability that a star
has a mass $M_\ast\ge 8\,M_\odot$, i.e., 
\be
\int_8^{120} f(m)dm = 0.005\;.
\ee
For completeness, we note that stars near the high end of the allowed mass
range can collapse directly to black holes (e.g., see the discussion
of \citealt{woos2017}). If black hole formation is efficient, then the
SLRs produced during stellar evolution will be incorporated into the
black holes instead of being distributed into the molecular cloud. 
Although the probability of black hole formation is not known, this
process would reduce the radioactive yields estimated in this 
paper, as illustrated below (see Figure 8).\footnote{For 
example, if all stars with $m>60$ collapse to black holes without
releasing SLRs, then the total radiocactive enrichment levels of 
this paper would be reduced by $\sim14\%$.}

Massive stars are placed at the center of their respective cluster 
and evolve into SN events upon reaching an age given by 
\be
\log_{10}\left({ \tau_{SN}\over {\rm Myr}}\right) = 
{1.4\over (\log_{10} m)^{1.5}}\;,
\ee
where this form represents an empirical fit \citep{williams2007} to
results from an ensemble of stellar evolution simulations
\citep{schaller1990}.

The locations of the remaining ``field" stars are then specified
within each cluster using a formalism similar to that used to place
clusters in the molecular cloud (with fractal dimension $d = 2$ for
our baseline model).  However, as stars are effectively point masses,
they can all be placed at the active sites of the final generation of
mature children, which is done randomly.  In addition, the lengths of
the cube in which stars are placed for a given cluster is scaled as
$r_{*c}=aR_c$ (with $a = 2$ for the baseline model), but active sites
that are farther than $r_{*c}$ from the cluster center are eliminated
as possible locations (thereby turning a cubic structure into a
spherical one).  We note that this process for placing field stars
does not accurately reflect the dynamics of cluster evolution, but
nevertheless, produces structures that are visually similar to those
observed in GMC environments.  The structure of a baseline GMC
generated using our scheme is shown in the left panel of Figure 1,
where blue spheres represent clusters (with radii $R_c$), black dots
show the location of massive stars, and grey spheres denote the
location of each field star.  For illustrative purposes, we also show
the corresponding column density map in the right panel of Figure 1,
where a Hernquist \citep{hernquist1990} density profile 
\be 
\rho(r) = {M_\infty\over 2\pi} 
{c\over r} {1\over (r+c)^3}\;\, 
\ee 
with $c = 0.2$ pc centered on each star is used to calculate the
density at each point of the region, and $M_\infty$ is set to $20\;
M_\odot$ so that the total mass of the GMC is $10^6\;M_\odot$
(ignoring any stellar mass contribution).  Noting that the mass
contained within a radius $r = c$ for a Henrnquist profile is
$M_\infty/4$, we define a characteristic density for star forming
regions as
\be 
\rho_* \equiv
{M_\infty /4 \over 4\pi c^3/3} = 
150 \, M_\odot \;{\rm pc^{-3} }\;.
\ee 
For ease of comparison to observed values of mass fractions in our
Solar System, the mass densities of $^{26}$Al and $^{60}$Fe obtained
in our work will be presented in terms of $\rho_*$.

\begin{figure}[t]
\begin{center}
\includegraphics[width=0.8\linewidth]{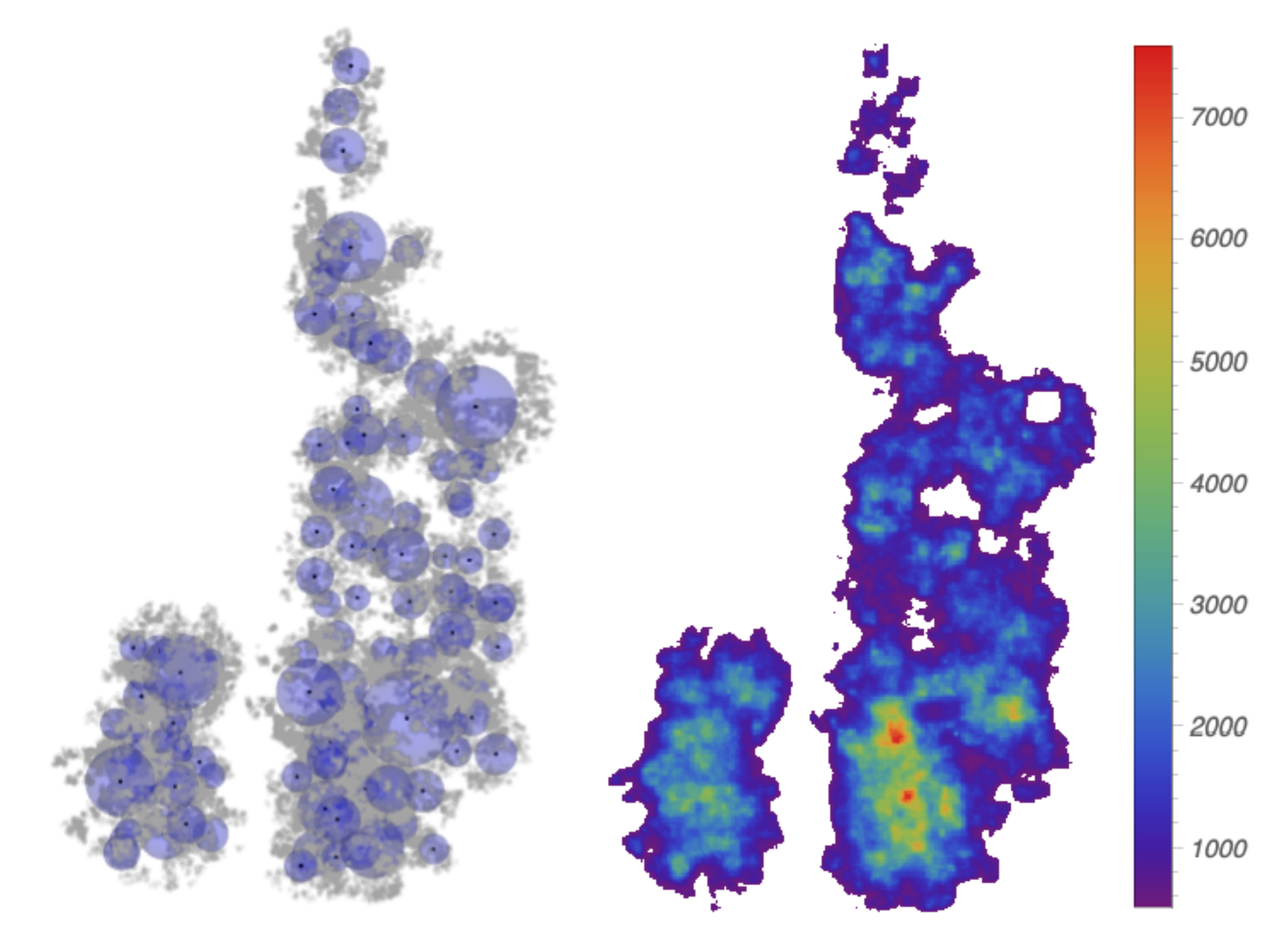}
\end{center}
\caption{{\bf Left:} A realization of a GMC generated randomly through
  the process described in \S 2.1 for the baseline model.  Blue
  spheres represent clusters (with radii $R_c$), black dots show the
  location of massive stars, and grey spheres denote the location of
  each field star.  {\bf Right:} Corresponding column density map for
  the GMC structure shown in the left panel, for which the gas density
  around each star follows a Hernquist profile (see Eq. (8) and
  corresponding text).  The bar legend is in units of $M_\odot / {\rm
    pc^2}$ }
\end{figure}

\subsection{Distribution of SLRs}

Massive stars ($M_\ast \ge 8\;M_\odot$) born in GMCs, while quite rare, 
live only a few million years before exiting in cataclysmic fashion. The
resulting supernova explosions that follow the death of those massive
stars spread the byproducts of the nuclear fusion that powers stellar
life, as well as nuclei created through the rapid capture of neutrons
during the SN event, into the surrounding giant molecular cloud
environment. The most massive of these stars ($M_\ast\ge 25\;M_\odot$) also
inject nucleosynthesis products into the surrounding medium via strong
winds during a short-lived Wolf-Rayet phase prior to their cataclysmic
ends.  Our main interest lies in the small fraction of those elements
that are unstable and will eventually undergo radioactive decay on
timescales $\sim 1 $ Myrs.  Specifically, $^{26}$Al, and $^{60}$Fe are
produced with relatively large abundances and have half-lives
$t_{1/2}$ of 0.72 and 2.6 million years, respectively -- long enough
for them to reach many of the stellar systems forming within the
molecular cloud, but short enough that their ensuing radioactive decay
can efficiently heat and ionize the disks within those systems.  For
completeness, we also consider the production of $^{26}$Al via stellar
winds during the Wolf-Rayet phase, as well as the production of the
lower-yield elements $^{36}$Cl ($t_{1/2}=0.30$ Myr) and $^{41}$Ca
($t_{1/2}=0.10$ Myr) in SN events.

The integrated abundance of $^{26}$Al produced during the Wolf-Rayet
phase of massive stars is presented in \cite{gounelle2012} for several 
different initial masses (see their Figure 3).   While the time
evolution of this process is complicated, the most important feature
for our study is the plateau reached just prior to the ensuing supernova 
event.  As a result, we adopt a highly idealized empirical model based 
on simple fits to the results of \cite{gounelle2012} wherein the
mass-dependent integrated abundance of $^{26}$Al has the form 
\be
M_{Al;w}(m)= 10^{-4.7+1.66(\log_{10}[m]-1.48)}\,M_\odot\;,
\ee
and remains constant over a time 
\be
\tau_w = \left[0.4+{1.6(m-30)\over 90}\right]\,{\rm Myr}\,,
\ee
prior to the SN event for stellar masses $M_\ast\ge 30\;M_\odot$.  The
abundance then diminishes due to radioactive decay following the SN
event.  Lower mass stars produce substantially lower yields, and are
therefore disregarded.  Our highly-idealized model is presented in
Figure 2, and can be compared directly to Figure 3 of 
\cite{gounelle2012}. 

\begin{figure}[t]
\begin{center}
\includegraphics[width=0.8\linewidth]{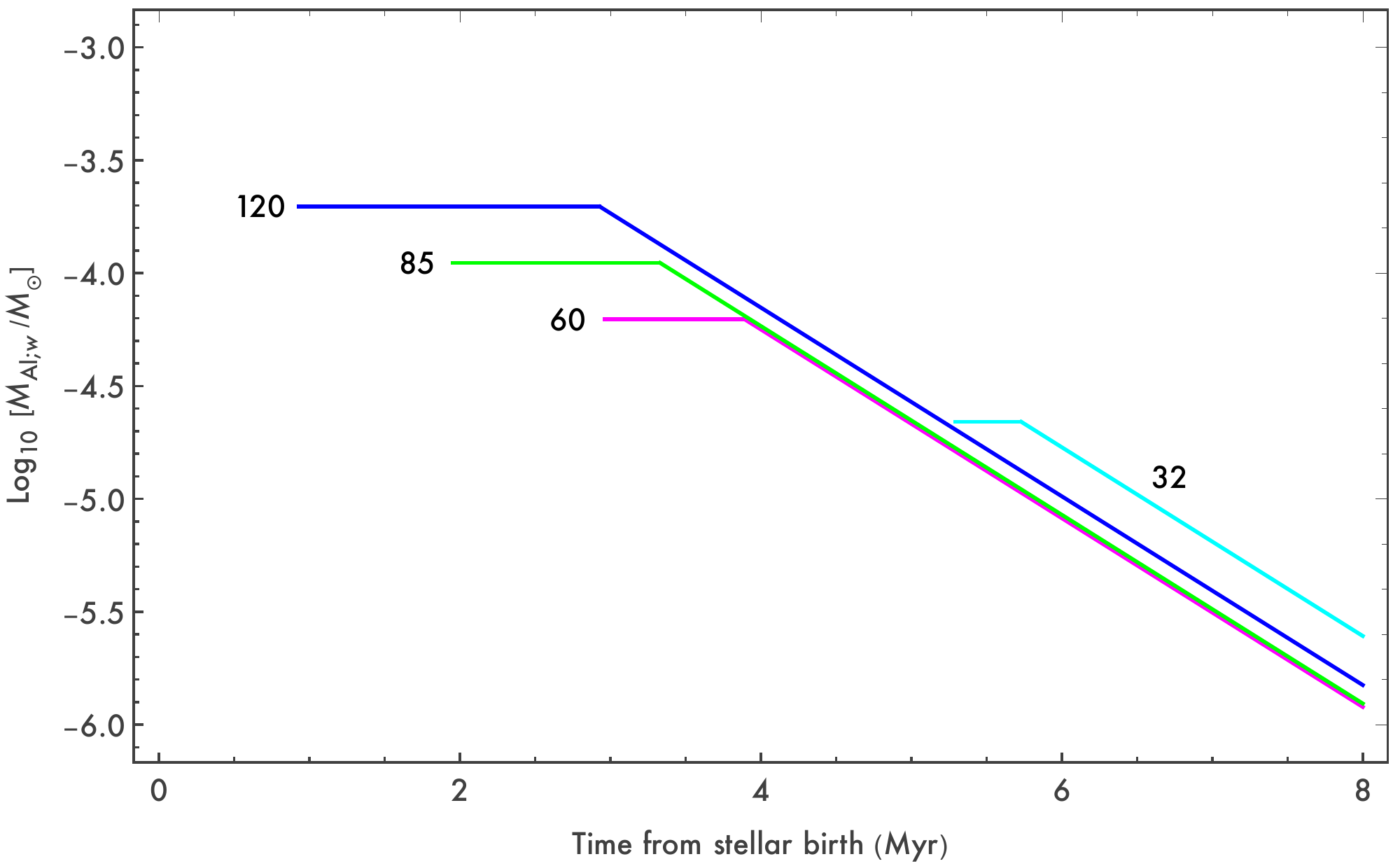}
\end{center}
\caption{Idealized empirical model for the integrated abundance of
  $^{26}$Al produced during the Wolf-Rayet phase of massive stars as a
  function of time from stellar birth for 4 initial progenitor masses,
  as indicated by the numbers next to the curves (in $M_\odot$).  The
  colors mimic those of Figure 3 in \cite{gounelle2012} on which our
  model is based.}
\end{figure}

The SN yields of SLRs used in this work are shown in Figure 3 over the
range of relevant stellar masses.  These yields are taken from the
stellar nucleosynthesis calculations of \cite{chieffi2013} (see also
\citealt{sukhbold2016,limongi2018}). These simulations provide updated
yields for the SLRs and include the effects of rotation (compare with
earlier stellar evolution models of \citealt{ww1995,lc2006}; see also
\citealt{timmes1995a,timmes1995b,woos2002,rauscher2002}).  The stellar
models are computed for discrete mass values, with the yields shown by
the solid circles, and with intermediate values determined by
interpolation (carried out in log-log space). Since results from the
stellar models are listed with a minimum mass of $m=13$, yields for
smaller progenitor masses are extrapolated assuming constant yields
over the range $8<m<13$. The resulting yields are shown in Figure 3
for the SLRs of interest: $^{26}$Al (blue curve), $^{60}$Fe (red
curve), $^{41}$Ca (green curve), and $^{36}$Cl (magenta curve). For
completeness, Figure 3 also includes the plateau $^{26}$Al
wind-generated abundances shown in Figure 2, along with the
corresponding fit from Eq. (10). 

\begin{figure}[t]
\begin{center}
\includegraphics[width=0.8\linewidth]{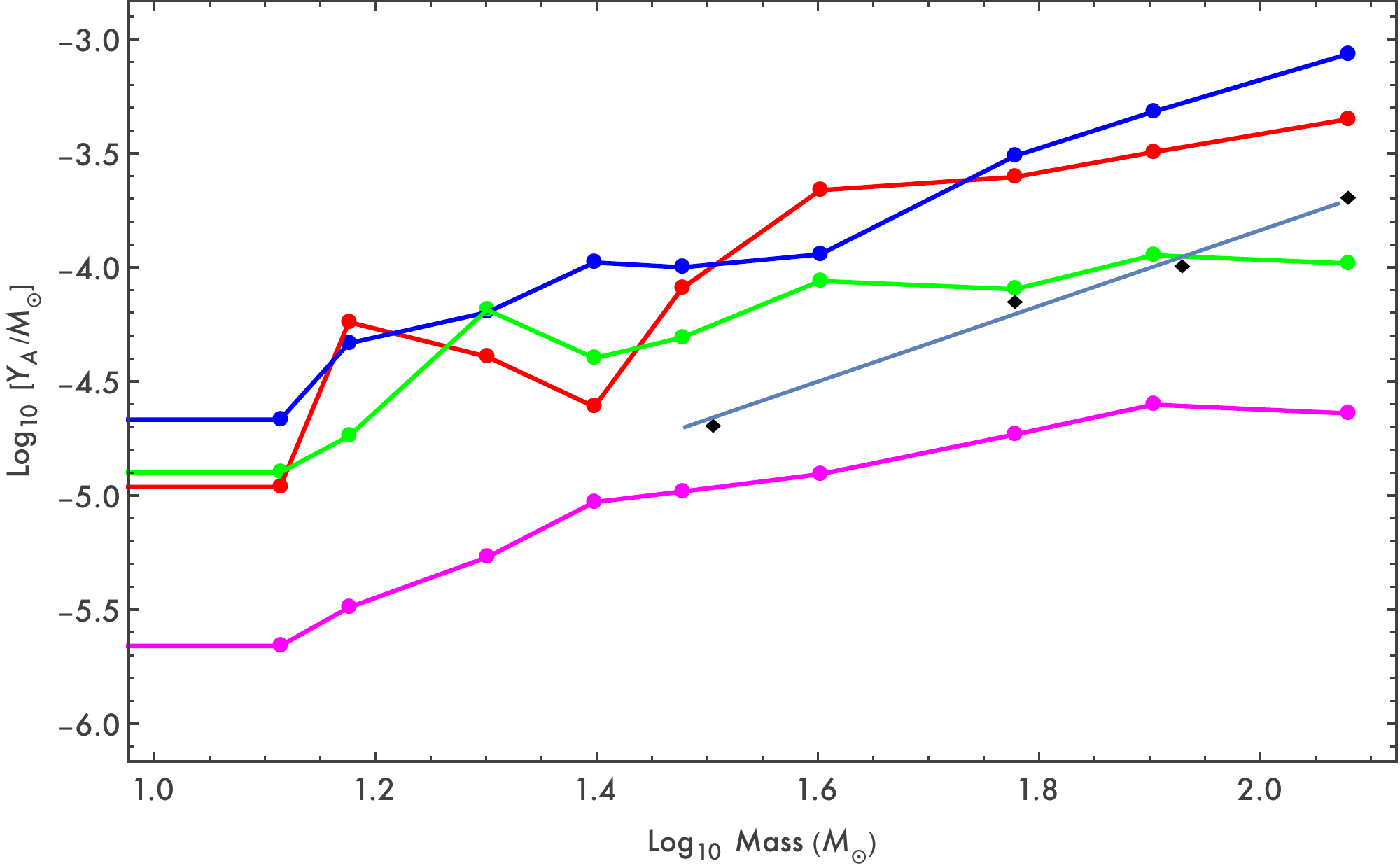}
\end{center}
\caption{Yields $Y_A$ of short-lived radioactive nuclei $^{26}$Al
  (blue), $^{60}$Fe (red), $^{41}$Ca (green), and $^{36}$Cl (magenta)
  as a function of progenitor mass.  Yields are taken from the stellar
  evolution models of \cite{chieffi2013}.  The curves show the
  interpolations and extrapolations used to determine yields for any
  given mass.  For completeness, the plateau $^{26}$Al wind-generated
  abundances from Figure 2 are presented by black diamonds along with
  the corresponding fit from Eq. (10).}
\end{figure}

For a given realization of a GMC environment, the total mass
enrichment of a given element as a function of time resulting from all
SN events, each occurring at time $t_i$, is given by
\be
M_A(t)=\sum_{i=1}^{N_{SN}}\;H[t-t_i] \;Y_{A_i}\;e^{-\lambda_A{(t-t_i)}}\,,
\ee
where $A$ denotes the species, $\lambda_A$ is the corresponding decay
constant, and $H[x]$ is the well-known step function.  Each
realization, in turn, generates a value of $M_A(t)$ that is a member
of the parent population which statistically describes mass enrichment
for our assumed GMC scenario.  The expectation value of the
corresponding parent distribution can be calculated by integrating
over both the IMF and stellar birth functions
\be
\langle M_A(t) \rangle= N^*_{mc}  
\int_8^{120}   f(m)\;  
\int_0^t \;b(t_*)  H[t-t_{SN}(m)] \;
Y_{A}(m)\;e^{-\lambda_A{(t-t_{SN}(m))}}\;dt_*\;dm\,,
\ee
where $t_*$ represents the stellar birth time, $t_{SN}=\tau_{SN}+t_*$,
and $b(t_*)$ is given by Eq. (4) for our baseline model.  The variance
can also be calculated via the expression
\be
\sigma^2 (t) = { \langle M_A(t)^2 \rangle-\langle M_A(t) 
\rangle^2\over \langle N_{SN}(t)\rangle-1}\;,
\ee
where
\be
\langle M_A(t)^2 \rangle= \langle N_{SN}(t)\rangle \;N^*_{mc}  \int_8^{120}   f(m)\;  \int_0^t \;b(t_*)  H[t-t_{SN}(m)] \;\left[ Y_{A}(m)\;e^{-\lambda_A{(t-t_{SN}(m))}}\right]^2\;dt_*\;dm\,,
\ee
and
\be
\langle N_{SN}(t)\rangle=N^*_{mc}  \int_8^{120}   f(m)\;  
\int_0^t \;b(t_*)  H[t-t_{SN}(m)]\; dt_*\;dm\,.
\ee
Similar expressions can be obtained for the distribution of $^{26}$Al
yields produced by winds, for which a given realization of a GMC
environment generates a total mass enrichment of
\be
M_{Al;w}(t)=\sum_{i=1}^{N_{SN}}\;H[t-t_i-\tau_w] 
H[t_i-t]\;M_{Al;w,i}+H[t-t_i] \;M_{Al;w,i}\;e^{-\lambda_A{(t-t_i)}}\,.
\ee

The expectation values calculated using the above integral expressions
are in complete agreement with mean value obtained from $10^5$
randomly generated realizations of GMCs for our baseline model (with
values from each realization calculated using Eq. [12]).  Likewise,
the variances calculated for the parent populations are in good
agreement with those obtained from our randomly generated sample,
though the latter are slightly wider owing to under sampling of the
IMF at high masses.  The mean values of the total mass enrichment
obtained from our numerical simulations of our baseline GMC model are
presented by the solid curves in Figure 4, with the corresponding
shading depicting the $\pm 1\;\sigma$ band.   For reference, the
mean values of the total mass enrichment at 5 Myr intervals of GMC
age, along with the corresponding characteristic relative abundance
for star forming regions, 
\be
\eta_A\equiv {\langle M_A \rangle/\langle V_{GMC}\rangle\over \rho_*}\,,
\ee
are compiled in Table 1, where the subscripts $Al$ and $Al+$ refer to
SN yields only, and SN plus wind yields, respectively.

\begin{table}
\begin{center}
\begin{tabular}{ccccccc}
\tableline\tableline
& 5\,{\rm Myrs}& 10\,{\rm Myrs} & 15\,{\rm Myrs} & 20\,{\rm Myrs}& 25\,{\rm Myrs} & 30\,{\rm Myrs} \\
\tableline
$\langle M_{Al} \rangle $ & $1.9\times 10^{-4}$ &$9.4\times 10^{-4}$ &$1.0\times 10^{-3}$ &$3.9\times 10^{-4}$ &$1.7\times 10^{-4}$  &$1.3\times 10^{-4}$  \\
$\langle M_{Al+}  \rangle$ & $3.7\times 10^{-4}$ &$1.3\times 10^{-3}$ &$1.2\times 10^{-3}$ &$3.9\times 10^{-4}$ &$1.7\times 10^{-4}$  &$1.3\times 10^{-4}$  \\
$\langle M_{Fe}  \rangle$ & $1.7\times 10^{-4}$ &$1.9\times 10^{-3}$ &$2.9\times 10^{-3}$ &$1.7\times 10^{-3}$ &$8.3\times 10^{-4}$  &$4.0\times 10^{-4}$  \\
$\langle M_{Ca}  \rangle$ & $1.1\times 10^{-5}$ &$5.6\times 10^{-5}$ &$6.5\times 10^{-5}$ &$2.6\times 10^{-5}$ &$1.2\times 10^{-5}$  &$9.7\times 10^{-6}$  \\
$\langle M_{Cl}  \rangle$ & $5.6\times 10^{-6}$ &$3.0\times 10^{-5}$ &$3.2\times 10^{-5}$ &$1.2\times 10^{-5}$ &$6.5\times 10^{-6}$  &$5.2\times 10^{-6}$  \\
$\eta_{Al} $ & $6.4\times 10^{-11}$ &$3.1\times 10^{-10}$ &$3.5\times 10^{-10}$ &$1.3\times 10^{-10}$ &$5.8\times 10^{-11}$  &$4.2\times 10^{-11}$  \\
$\eta_{Al+} $ & $1.2\times 10^{-10}$ &$4.2\times 10^{-10}$ &$4.0\times 10^{-10}$ &$1.3\times 10^{-10}$ &$5.8\times 10^{-11}$  &$4.2\times 10^{-11}$  \\
$\eta_{Fe} $ & $5.8\times 10^{-11}$ &$6.4\times 10^{-10}$ &$9.5\times 10^{-10}$ &$5.7\times 10^{-10}$ &$2.8\times 10^{-10}$  &$1.3\times 10^{-10}$  \\
$\eta_{Ca} $ & $3.7\times 10^{-12}$ &$1.9\times 10^{-11}$ &$2.2\times 10^{-11}$ &$8.8\times 10^{-12}$ &$4.0\times 10^{-12}$  &$3.2\times 10^{-12}$  \\
$\eta_{Cl} $ & $1.9\times 10^{-12}$ &$1.0\times 10^{-11}$ &$1.1\times 10^{-11}$ &$3.9\times 10^{-12}$ &$2.2\times 10^{-12}$  &$1.7\times 10^{-12}$  \\
 \tableline
\end{tabular}
\caption{Mean values of the total mass enrichment from $10^5$
  realizations of our baseline GMC model at 6 different GMC ages,
  along with the corresponding characteristic relative abundance
  $\eta$ for star forming regions (as defined in Eq. [18]).  All
  masses are given in units of $M_\odot$}
\end{center}
\end{table}

Further insight is provided by the skewness and kurtosis of the
generated $^{26}$Al and $^{60}$Fe distributions, which are plotted
along with the normalized first central moments $\langle
M_A\rangle/\sigma$ as a function of GMC age in Figure 5. Consistent
with the results shown in Figure 4, the first central moments (solid
curves) increase with age, indicating that the distributions narrow
with age as the death of more stars increases the sample size of SN
events which contribute to the yields.  The skewness and kurtosis of
the distributions decrease with age from fairly large initial values
to values of around 0.5 and 3.1, respectively, both of which are
slightly larger than their normal counterparts of 0 and 3.  These
results indicate that the distributions evolve from highly
non-Gaussian forms to nearly normal distributions with a median value 
given by $\sim \langle M_A\rangle -\sigma/6$. Moreover, the tails are
slightly heavier and longer at higher mass-enrichment values than
normal distributions.  To illustrate this point, we show in Figure 6
the histogram of the $^{26}$Al mass-enrichment values generated at a
GMC age of 30 Myrs that correspond to the results shown in Figures 4
and 5, along with a Gaussian fit to the data.  Finally, the
probability functions for $\log[M_{Al}]$ (with and without wind
contributions) and $\log[M_{Fe}]$ resulting from our analysis are
shown at 5 Myr intervals of GMC age in Figure 7.  While the means of
the $^{26}$Al and $^{60}$Fe distributions reach their highest values
at $t \approx 15$ Myrs when only SN events are considered, the
contribution from winds maximizes the mass-enrichment of $^{26}$Al at
$t \approx 10$ Myrs.

The discussion thus far assumes that stellar progenitors of all 
masses contribute to the production of SLRs. In some cases, however,
supernova explosions can stall, so that the stars directly collapse to
form black holes. In such cases, the SLRs that are produced via
stellar nucleosynthesis are not available for distribution throughout
the cloud. Unfortunately, the masses for which black hole production
occurs is not well determined. Simulations of supernova detonation
show complicated behavior, with some progenitor masses producing black
holes and comparable masses having successful explosions. Moreover,
the trends are not monotonic and different sets of simulations are
not in complete agreement (e.g., compare
\citealt{sukhbold2016,couch2020,boccioli2021}).  Nonetheless, the
likelihood of black hole formation generally increases with progenitor
mass.  In addition, gravity wave experiments
\citep{abbott2016,abbott2021} are finding black holes in the mass
range $20-120M_\odot$.  After correcting for observational biases
(e.g., \citealt{croker}), the underlying mass distribution of black
holes dispalys a broad peak at $M_{bh}\approx60M_\odot$. The
corresponding distribution of progenitor masses (for stars that
produce black holes) is expected to peak at somewhat higher mass.

In order to assess the loss of SLRs due to black hole formation, 
we calculate the total SNR mass enrichment for $^{28}$Al and
$^{60}$Fe in our baseline model, but with the contributions from stars
with mass $M\ge$ 100, 80, and 60 $M_\odot$ removed.  The results are
shown in Figure 8, along with the curves corresponding to the full
mass range and corresponding $\pm\sigma$ boundaries shown in Figure 4.
We note that the formation of black holes could have a noticeable
effect on our results for GMC ages less than 15 Myrs due primarily to
the delay in the onset of SLR injection (we note that a 60 $M_\odot$
star remains on the main sequence for $\sim 1$ Myr longer than a 120
$M_\odot$ star), but also because the total contribution of SLRs from
stars of mass $\ge 60 M_\odot$ is about 14\% of the total.
Nevertheless, as can be seen by comparing the curves in Figure 8 to
the $\pm\sigma$ boundaries of the full mass range contribution, the
effect is comparable to (or less than) the statistical fluctuations
that arise naturally in the system.

\begin{figure}[t]
\begin{center}
\includegraphics[width=0.8\linewidth]{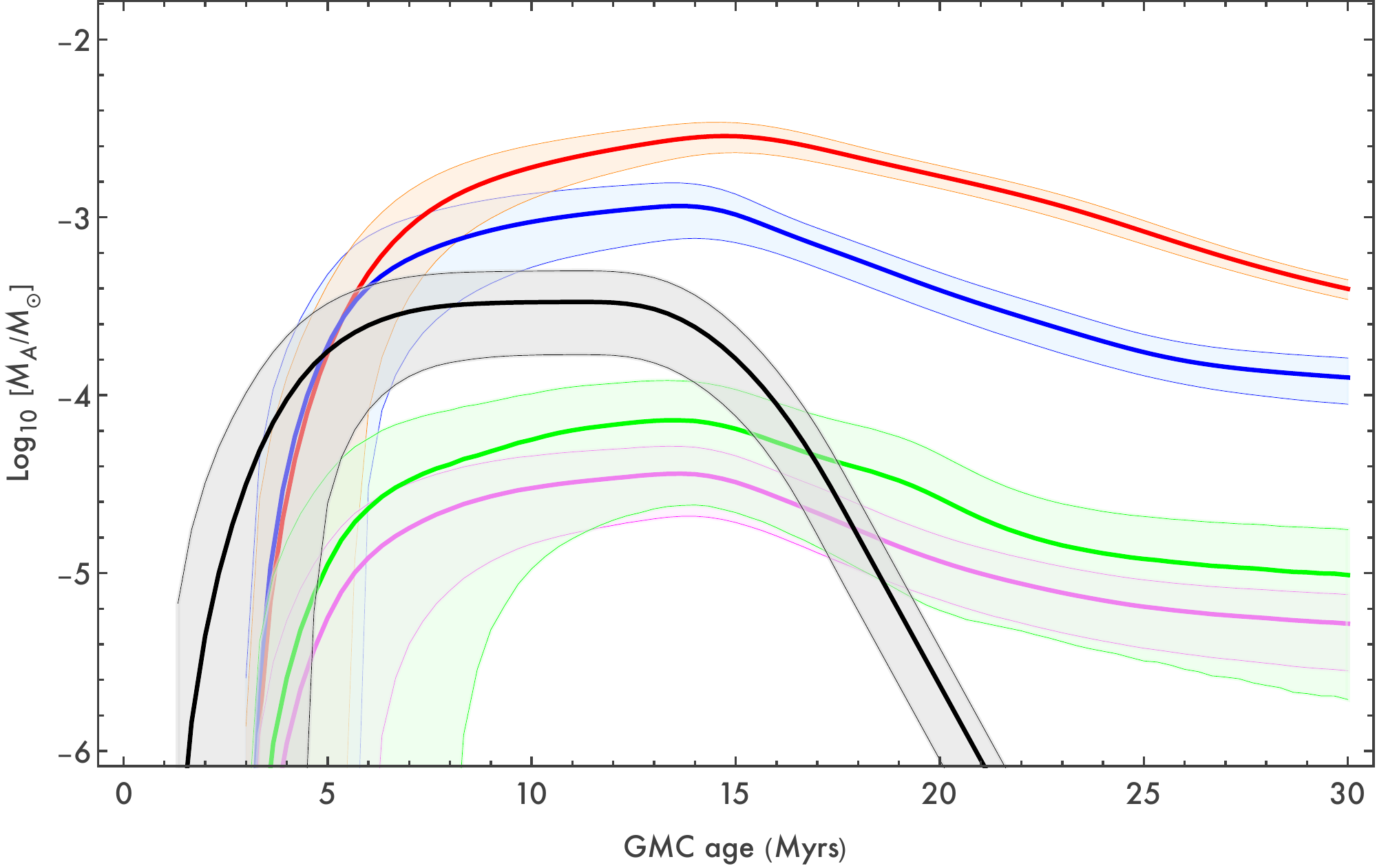}
\end{center}
\caption{The solid curves depict the mean values of the distributions
  of total mass enrichment resulting from all SN events in a GMC
  obtained from $10^5$ realizations of our baseline model.  The
  shading around the solid curves represent $\pm \sigma$ boundaries of
  the corresponding distributions.  The color scheme
  is the same as same as in Figure 3: $^{60}$Fe (red); $^{26}$Al from
  SN events only (blue); $^{41}$Ca (green); $^{36}$Cl (magenta);
  $^{26}$Al from Wolf-Rayet winds (black).  }
\end{figure}

\begin{figure}[t]
\begin{center}
\includegraphics[width=0.8\linewidth]{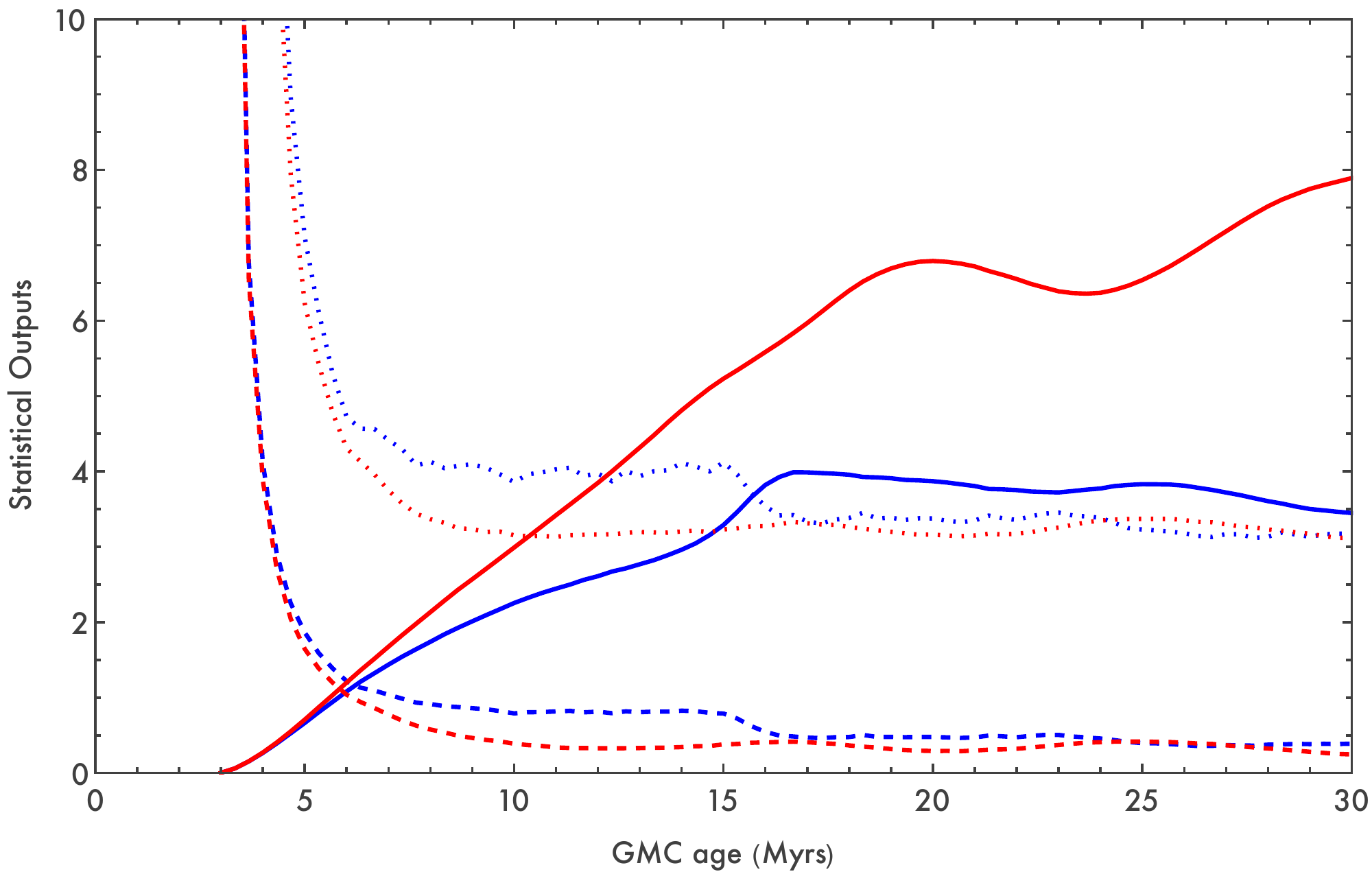}
\end{center}
\caption{Normalized central moments as a function of GMC age of the
  $^{60}$Fe (red) and $^{26}$Al (blue) distributions of total mass
  enrichment as a function of GMC age obtained from $10^5$
  realizations of our baseline model.  Solid curves depict the first
  normalized central moment $\langle M_A\rangle/\sigma$, the dashed
  curves depict the skewness, and the dotted curves depict the
  kurtosis.  }
\end{figure}

\begin{figure}[t]
\begin{center}
\includegraphics[width=0.8\linewidth]{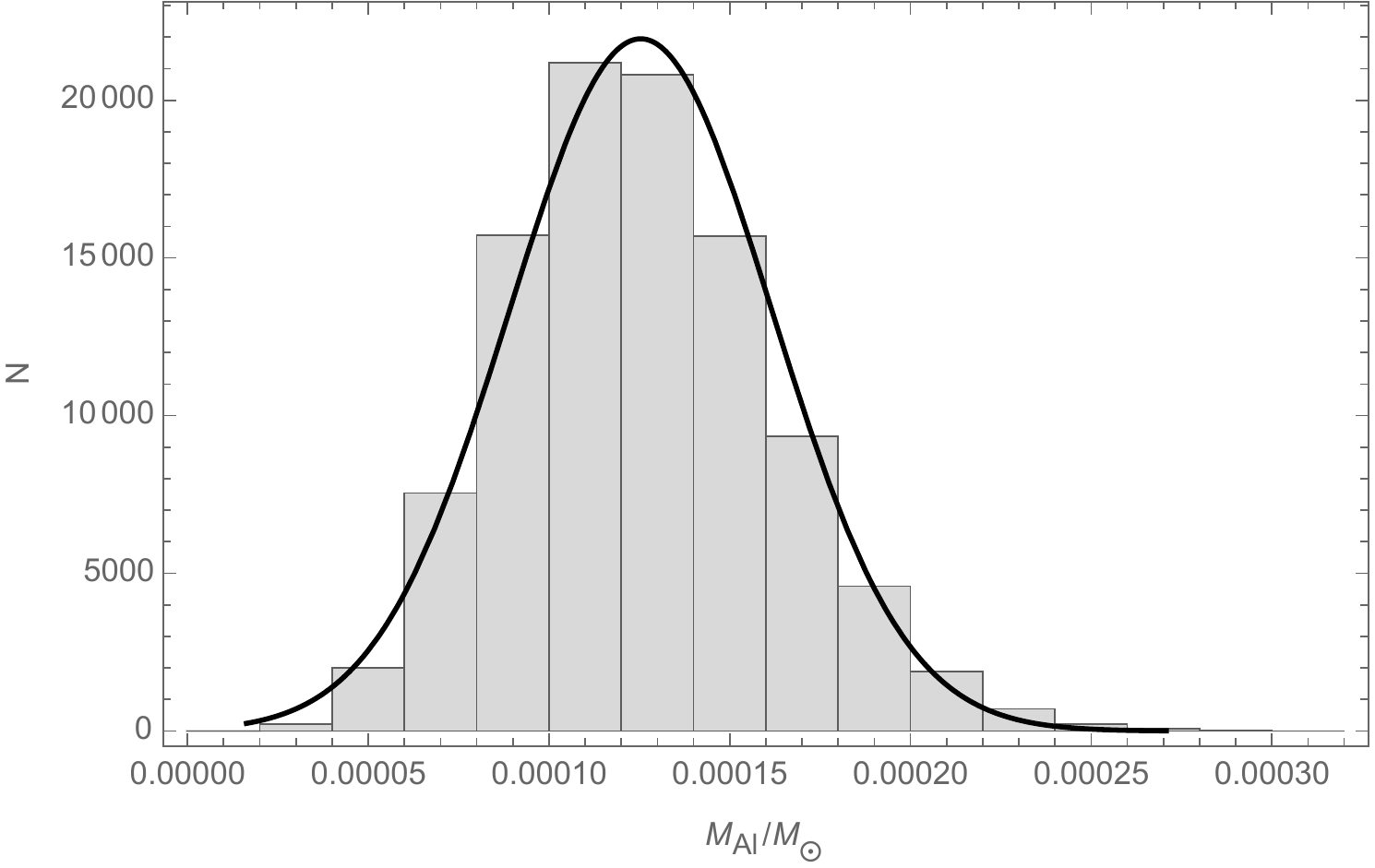}
\end{center}
\caption{Histogram of the values of total mass enrichment of $^{26}$Al
  in a GMC of age 30 Myr obtained from $10^5$ realizations of our
  baseline model, along with the corresponding normal distribution
  with the same mean and standard deviation as the resulting sample.}
\end{figure}

\begin{figure}[t]
\begin{center}
\includegraphics[width=0.8\linewidth]{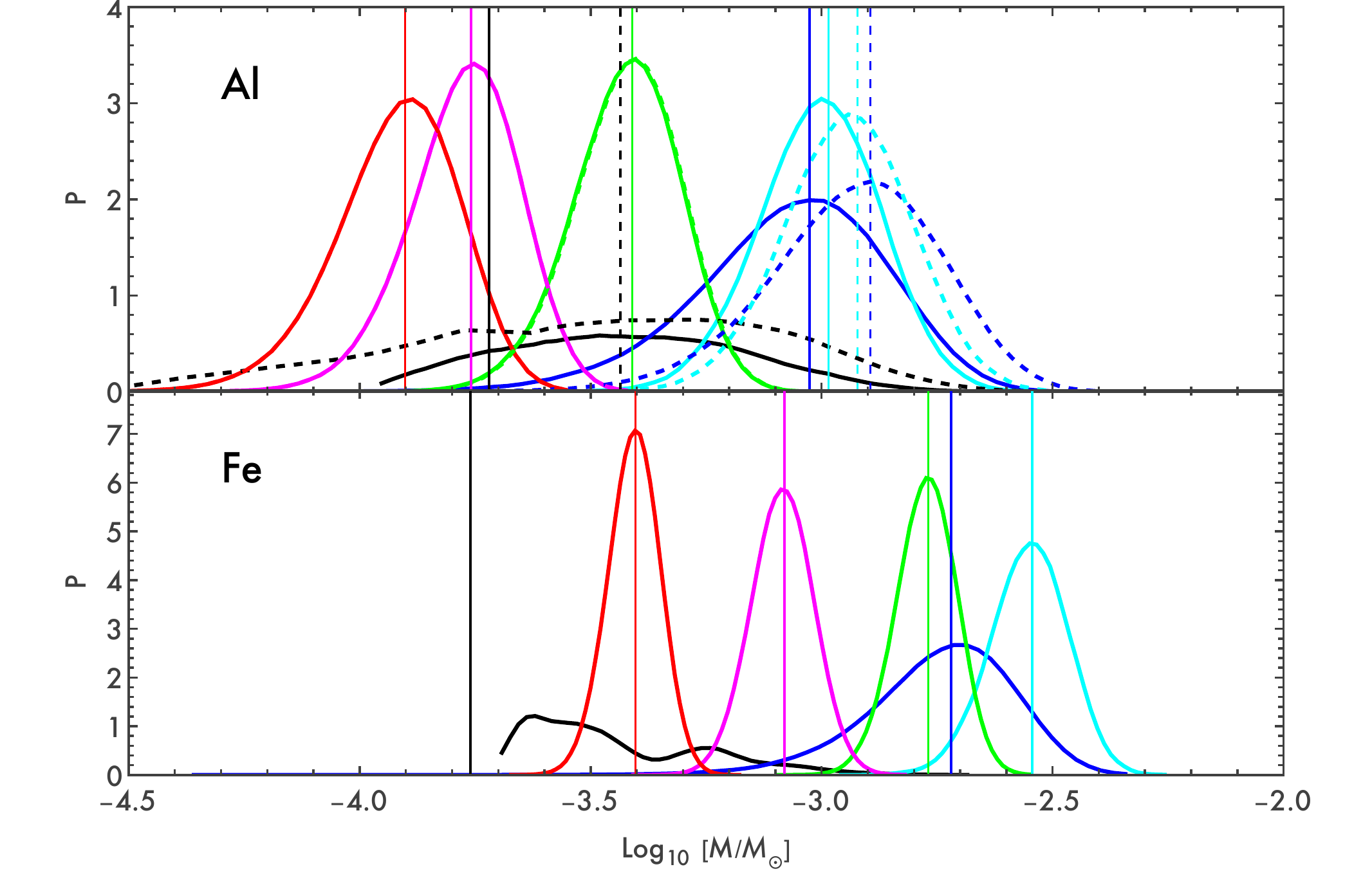}
\end{center}
\caption{Probability functions for the values of the total GMC mass
  enrichment (in units of $M_\odot$) for $^{26}$Al (top panel) and
  $^{60}$Fe (bottom panel). These results were obtained from $10^5$
  realizations of our baseline model at time $t$ = 5 (black), 10
  (blue), 15 (light blue), 20 (green), 25 (magenta) and 30 (red) Myr.
  Solid curves denote enrichment from SN events only, whereas the
  dashed curves in the top panel include $^{26}$Al produced by the
  winds of massive stars during their Wolf-Rayet phase.  The vertical
  lines denote the mean values of the corresponding mass
  enrichment distributions (e.g., $\log\langle M_A\rangle$).  Note
  that no SN occurred in many realizations of our model at 5 Myr, and
  these results, while not shown in the figure, are included in the
  calculation of the mean value for the corresponding distributions.}
\end{figure}

\begin{figure}[t]
\begin{center}
\includegraphics[width=0.8\linewidth]{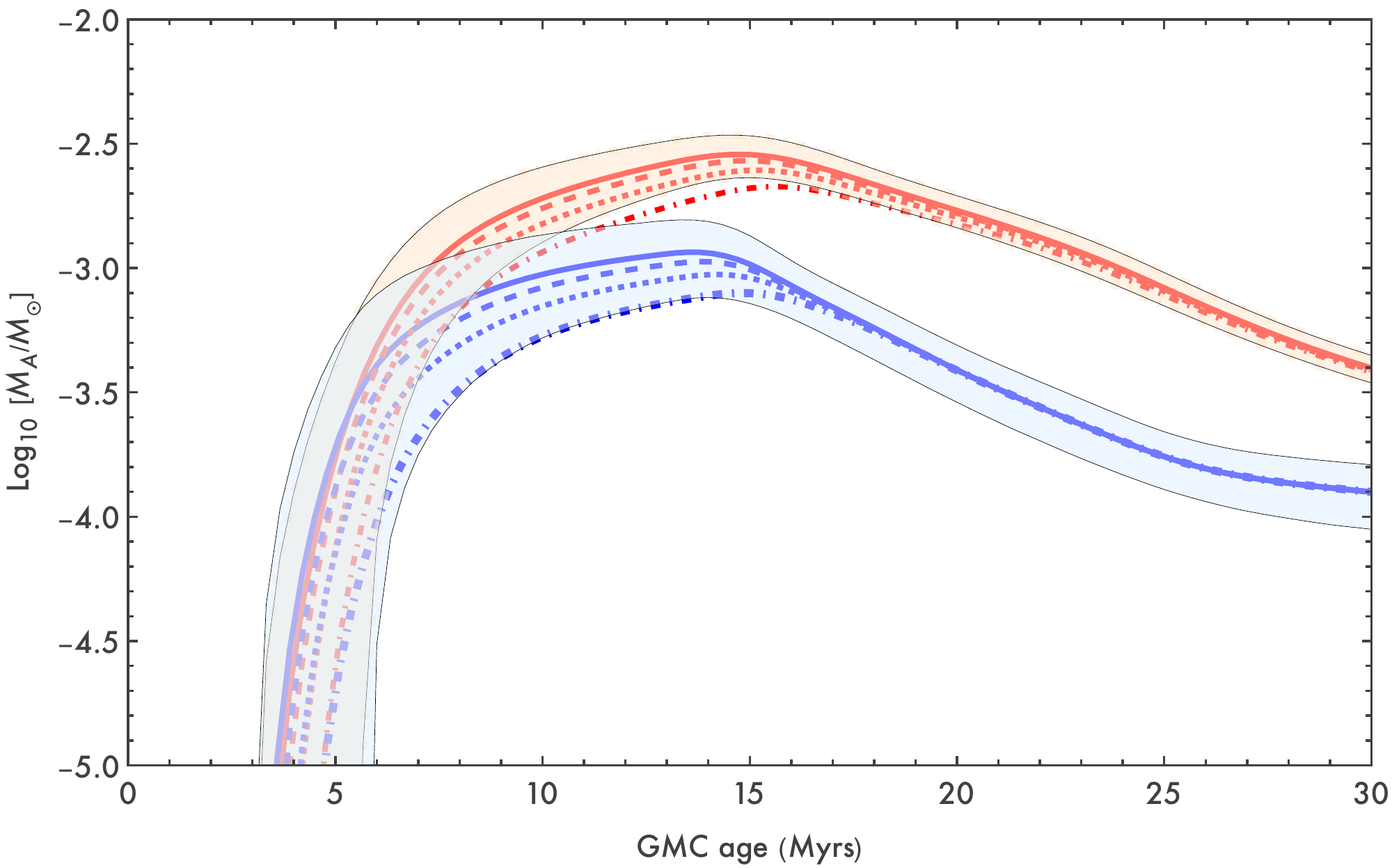}
\end{center}
\caption{The solid curves depict the mean values of the distributions
  of total mass enrichment resulting from all SN events in a GMC
  obtained from $10^5$ realizations of our baseline model, as shown in Figure 4.  The
  shading around the solid curves represent $\pm \sigma$ boundaries of
  the corresponding distributions ($^{60}$Fe  - red; $^{26}$Al from
  SN events only - blue).  The remaining curves depict the same
  values, but with contributions from stars with masses 
  $\ge 100 M_\odot$ (long-dashed), $80 M_\odot$ (short-dashed), and
  $60 M_\odot$ (dot-dashed) excluded. }
\end{figure}

\subsection{Relative Abundances of SLR's in Star Formation Sites}

Determining the distributions of the relative abundances of SLRs
in the dense regions where stars form requires an understanding of how
radioactive nuclei propagate through the GMC environments. 
Detailed numerical simulations of supernova
remnants indicate that the blast wave and ejecta propagate freely for
distances of order 1 -- 2 pc and then interact strongly with the
background molecular cloud (e.g., \citealt{pan2012}). The ejecta are
efficiently mixed into the cloud within a relatively thin layer,
with thickness $\sim0.5$ pc, and within a time frame of $\sim30,000$
years. Subsequent movement of the radioactive nuclei
must take place diffusively.\footnote{Note that in the limit where the 
diffusion constant becomes large, the propagation of the SLRs becomes 
effectively free-streaming. As a result, the diffusive model includes 
the free-streaming limit.} The nuclei themselves are tied to the
magnetic fields of the cloud, but the gyroradius is small compared to
the size scales of interest. As a result, the subsequent propagation
of the SLRs occurs primarily through turbulent diffusion. 

Since the mixing size and time scales are smaller than those of
interest for the propagation of SLRs throughout the cloud, we assume
here that each supernova immediately delivers its SLRs to a thin shell
with radius $R = 2$ pc, and calculate the ensuing density profiles
(including decay) for both $^{26}$Al and $^{60}$Fe relative to the
shell center by assuming point-diffusion from each part of the shell.
Doing so yields an expression 
\be
\rho(r, t)={2 M_A D  t \;e^{-( r^2+R^2)/(4D t)}\;e^{-\lambda_A t}\;
\over  rR[4\pi D t]^{3/2}}\sinh\left[{2 rR\over 4D t}\right]\,,
\ee
for the density of the diffusing particles at radius $r$ from the SN event at time $t$ measured relative to that event.
The diffusion constant (see 
\citealt{klessen2003}) is estimated through the expression  
\be
D = 2 v_T \ell \,,
\label{diffcon} 
\ee
where $v_T$ is the turbulent transport speed and $\ell$ is the mean
free path. The speed $v_T$ is identified with the observed non-thermal
linewidths $(\Delta v)$ of the molecular clouds, where $(\Delta v)$
scales with the size $L$ of the observed region according to 
\be
v_R \sim (\Delta v) \sim 1 {\rm km/s} 
\left({L \over1\,{\rm pc}}\right)^b\,,
\label{vscale} 
\ee
where the index $b=0.4-0.5$ \citep{larson1981,jijina1999}.  Equation
(\ref{vscale}) holds over a range of size scales $L=1-30$ pc, with a
corresponding range of velocity scales $\sim1-5$ km/s. These values
imply that the effective diffusion constant lies in the range
$D\sim1-150$ (km/s)pc. Given the uncertainty within this possible
range, we adopt $D=10$ (km/s)pc for our baseline model (roughly, the
geometric mean of the range), but will also explore how our results
vary for other values.

In contrast to the SN ejecta, the $^{26}$Al produced during the
Wolf-Rayet phase of a massive star is carried by stellar winds out to
the boundary where shocks stall the winds.  Given that wind speeds are
typically $\sim 100$ km/s, we assume that this process instantaneously
distributes the produced $^{26}$Al uniformly within a sphere of radius
$R_w = 5$ pc centered on the massive star 
\citep{gounelle2012,deharveng2010}. 

For a given realization of our baseline model for a GMC, the densities
$\rho_{Al}$ and $\rho_{Fe}$ at every field star location are
determined at a given GMC age by summing over the contributions of all
SN events.  The process is then repeated for a total of $10^4$ GMC
realizations (thereby building up a total sample of $\approx 5\times
10^8$ field point locations). Corresponding values for the relative
abundances of SLRs are obtained by dividing the calculated densities
by the characteristic density of a star forming region $\rho_* = 150
\,{\rm M_\odot\, pc^{-3}}$ (see Eq. [9]).  The resulting probability
functions for $\log[\rho_{Al}/\rho_*]$ (with and without wind
contributions) and $\log[\rho_{Fe}/\rho_*]$ at 5 Myr intervals of GMC
age are shown in Figure 9.  The mean values for each distribution are
indicated by the corresponding long vertical lines.  For
comparison, the characteristic relative abundances presented in Table
1 are shown by the short vertical lines at the top of the figure.

The distributions depicted in Figure 9 show a number of trends. Early
on, SLRs injected into the GMC have not had much time to propagate, 
resulting in large spatial density fluctuations characterized by broad
distributions with mean values larger than the expected characteristic
values.  Over time, diffusion smooths out the fluctuations, and the
relative abundance distributions become somewhat narrower. Taken as a
whole, the distributions of SLRs span several orders of magnitude,
with the bulk of the results falling in the range $10^{-11}-10^{-8}$
for both radioactive species of interest (where this range is
consistent with previous estimates -- see \citealt{vasileiadis2013},
\citealt{kuffmeier2016}). Significantly, the high end of the range
includes Solar System benchmark values ($X_{26}\sim4\times10^{-9}$ and
$X_{60}\sim10^{-9}$), although the typical mass fractions are somewhat 
lower. 

In addition, we note that diffusion competes with radioactive decay to
remove SLRs from the GMC environment with a characteristic time scale
that depends on the diffusion constant, 
\be
\tau_D = {R_{GMC}^2\over 4 D}=7\,{\rm Myr}\;
\left({R_{GMC}\over 17\,{\rm pc}}\right)^2\;
\left({D\over 10\,{\rm (km/s)pc}}\right)^{-1}\,,
\ee
and which is comparable to the GMC age for our baseline model. Since
$\tau_D$ is larger than the half-lives of our SLRs, diffusion is not
expected to have a significant effect on lowering the SLR content for
the GMC for our standard scenario (notice, however, that we consider
more efficient SLR propagation out of the cloud in the following
section). We also note that the difference between the distribution
means and expected characteristic values become more pronounced at
later ages for the longer-living $^{60}$Fe.

\begin{figure}[t]
\begin{center}
\includegraphics[width=0.8\linewidth]{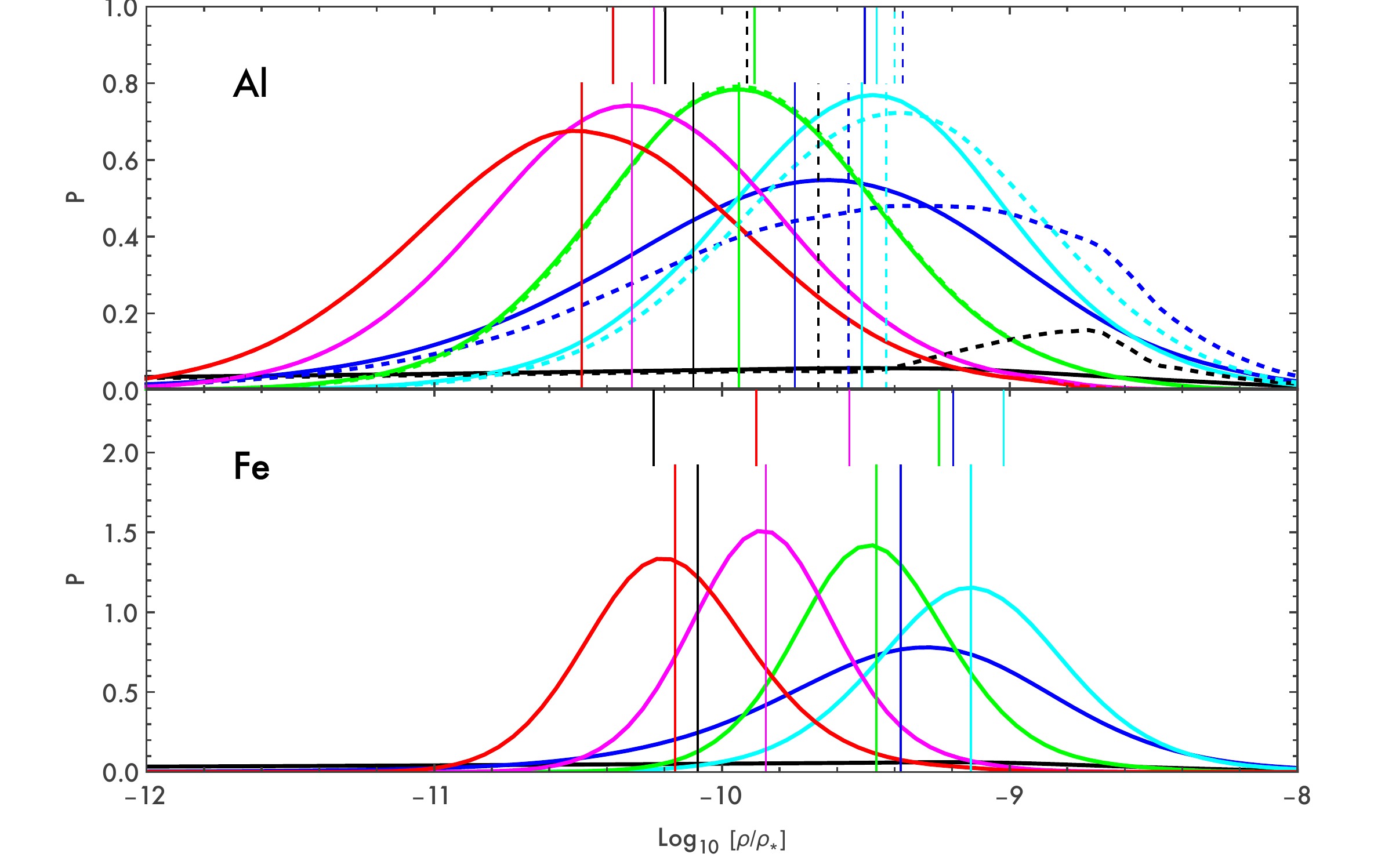}
\end{center}
\caption{Baseline model probability functions for the logarithmic
  values of SLR densities at stellar sites for $^{26}$Al (top panel --
  dashed curves include wind) and $^{60}$Fe (lower panel) at 5
  (black), 10 (blue), 15 (light blue), 20 (green), 25 (magenta) and 30
  (red) Myrs.  The lower vertical lines denote the mean value of the
  corresponding distribution.  The upper solid vertical lines show the
  corresponding values of $\log[\langle M_A\rangle /\langle
    V_{GMC}\rangle]$.  All densities are in units of the mean stellar
  site density $\rho_* = 150\;M_\odot \;{\rm pc^{-3}}$.  Note that no
  SN occurred in many realizations of our model at 5 Myr, and these
  results, while not shown in the figure, are included in the
  calculation of the mean value for the corresponding distributions.}
\end{figure}

In addition to the absolute values of the radioactive abundances, it
is also interesting to consider the ratio of different SLRs.  We
complete our analysis by plotting the probability distributions for
the abundance ratios $\log[\rho_{Fe}/\rho_{Al}]$ (with and without
winds) calculated in the star forming regions at our six
representative ages in Figure 10. The corresponding ratios for
$\log[\rho_{Cl}/\rho_{Al}]$ and $\log[\rho_{Ca}/\rho_{Al}]$ are shown
in Figure 11.  

\begin{figure}[t]
\begin{center}
\includegraphics[width=0.8\linewidth]{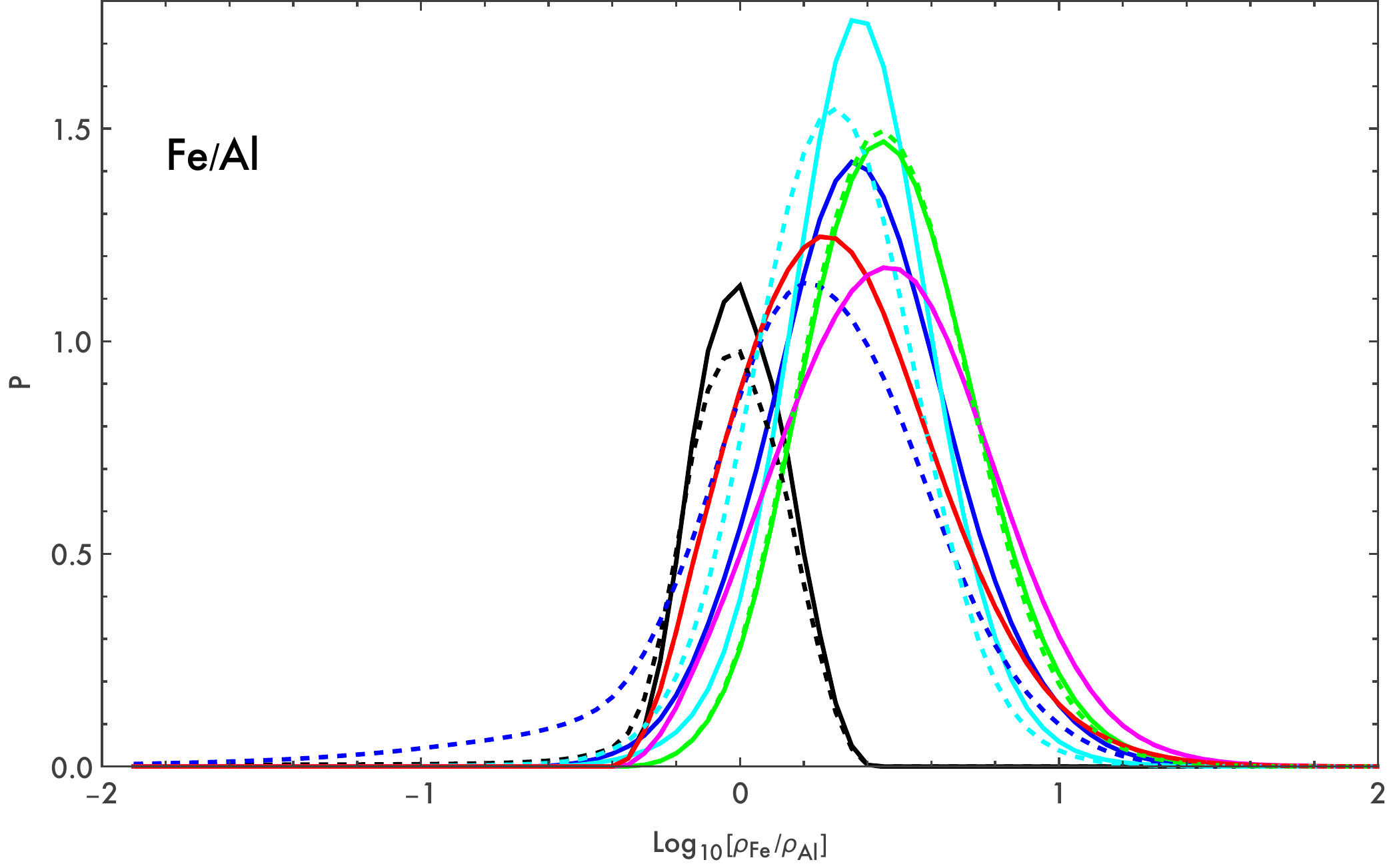}
\end{center}
\caption{Baseline model probability functions for the logarithmic
  values of the $^{60}$Fe to $^{26}$Al density ratios at stellar sites
  at 5 (black), 10 (blue), 15 (light blue), 20 (green), 25 (magenta)
  and 30 (red) Myrs.  Dashed curves includes wind contributions to
  $^{26}$Al.  }
\bigskip 
\end{figure}

The distributions of SLR abundance ratios for $^{60}$Fe/$^{26}$Al
shown in Figure 10 display a peak at $\sim2.5$ and extend down to
$\sim0.3$. This lower value is roughly comparable to the benchmark
value $X_{60}/X_{26}\approx0.25$ that we would expect for the early
Solar System, if $^{60}$Fe enrichment occurs at levels corresponding
to the upper end of the measured range. For comparison, the abundance
ratio inferred from gamma-ray observations \citep{diehl2006,diehl2013}
is $\sim0.6$, which corresponds to a galactic scale and steady-state
average. Although this value falls within the expected range, it is
below both the peak and the mean of the distribution. We also note
that these distributions are generally above the upper limits used in
many considerations of the nascent Solar System \cite{gounelle2015}.
One possible interpretation of this finding is that another mechanism
(e.g., spallation) could have contributed to the production of
$^{26}$Al when our Solar System formed, thereby lowering the ratio.

\begin{figure}[t]
\begin{center}
\includegraphics[width=0.8\linewidth]{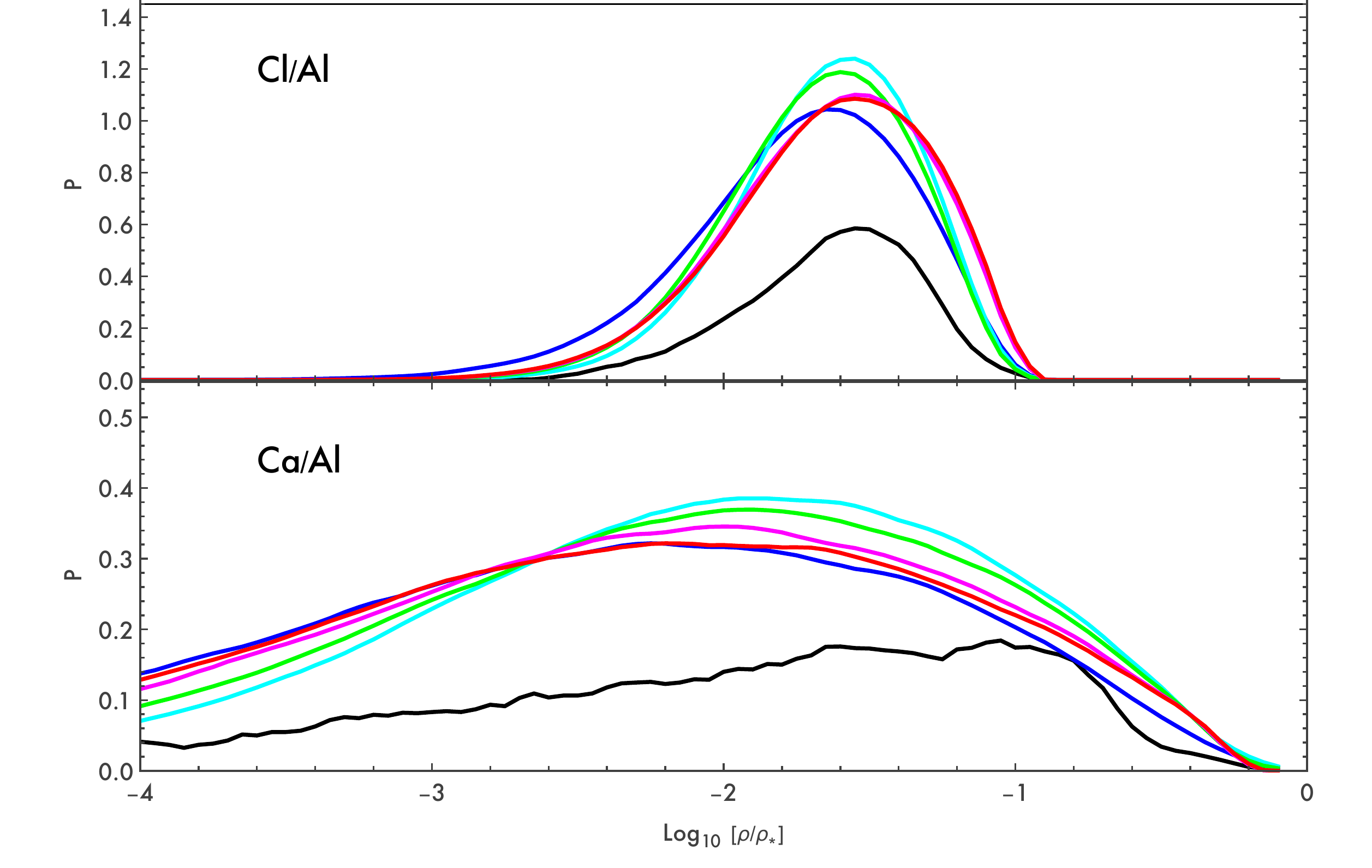}
\end{center}
\caption{Baseline model probability functions for the logarithmic
  values of the $^{36}$Cl to $^{26}$Al density ratios (top panel) and
  $^{41}$Ca to $^{26}$Al density ratios (bottom panel) at stellar
  sites at 5 (black), 10 (blue), 15 (light blue), 20 (green), 25
  (magenta) and 30 (red) Myrs.  Wind contributions to $^{26}$Al are
  not included. }
\end{figure}

\section{Analysis of Different Enrichment Scenarios}
\label{sec:variations} 

In this section we perform a sensitivity analysis of our nuclear
enrichment model by considering several variations of the input
parameters. Specifically, we calculate the $^{26}$Al and $^{60}$Fe
abundance distributions at star formation sites for different star
formation scenarios, varying physical parameters that determine
molecular cloud structure, and different properties for the transport
of radioactive material through the cloud.  In all cases, the number
of stars is held fixed at $N_{mc}^* = 50,000$, and the GMC dimensions
are sampled as in our baseline model. The resulting distributions of 
SLR abundances can then be compared to the those shown in Figure 9.

\subsection{Instantaneous Star Formation}

One limiting case for the star formation history of a GMC is to have
all of the stars form at once.  In this scenario, all stars form at
$t=0$, so that the stellar birth distribution function is given by
$b(t)=\delta(t)$.  All other model parameters are the same as for our
baseline model.  As can be seen from Figure 12, the immediate onset of
star formation injects SLRs into the GMC at earlier times, leading to
the yields of both $^{60}$Fe and $^{26}$Al peaking at around $4 - 5$ Myrs
(compare with Figures 2 and 3 from \citealt{voss2009}).  

The resulting probability functions for $\log[\rho_{Al}/\rho_*]$ (with
and without wind contributions) and $\log[\rho_{Fe}/\rho_*]$ at 5 Myr
intervals of GMC age are shown in Figure 13.  The mean values for each
distribution are indicated by the corresponding lower vertical lines,
and can be compared directly to their baseline model counterparts,
which are shown by the upper vertical lines.  Not surprisingly,
instantaneous star formation produces higher SLR densities for the
first $10$ Myrs, and lower SLR densities at later times.

\begin{figure}[t]
\begin{center}
\includegraphics[width=0.8\linewidth]{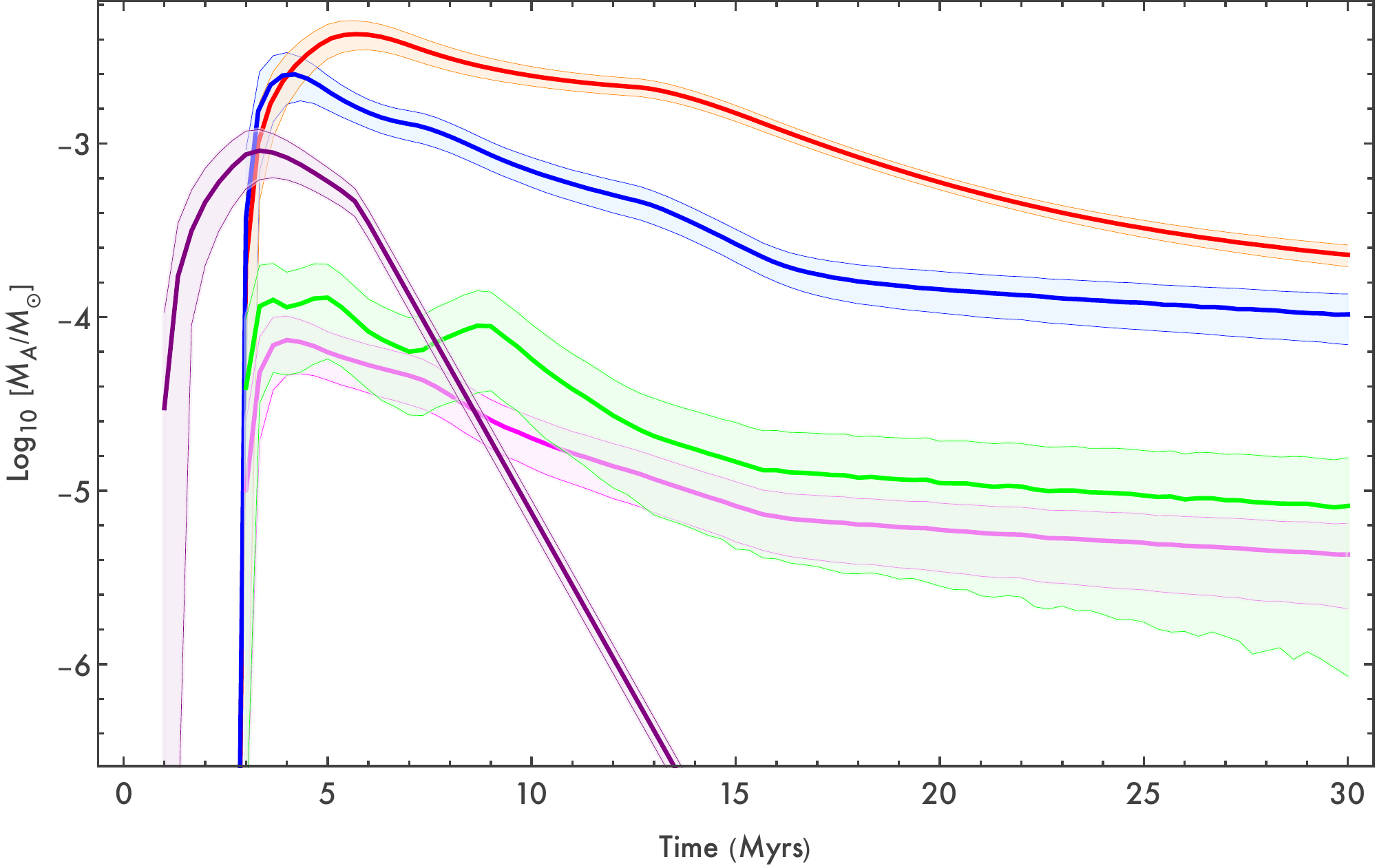}
\end{center}
\caption{The solid curves depict the mean values of the distributions
  of total mass enrichment resulting from all SN events in a GMC
  obtained from $10^5$ realizations of our instantaneous star formation
  model.  The shading around the solid curves represent $\pm \sigma$
  boundaries of the corresponding distributions. The color
  scheme is the same as same as in Figure 3: $^{60}$Fe (red);
  $^{26}$Al from SN events only (blue); $^{41}$Ca (green); $^{36}$Cl
  (magenta); $^{26}$Al from Wolf-Rayet winds (black).  }
\end{figure}

\begin{figure}[t]
\begin{center}
\includegraphics[width=0.8\linewidth]{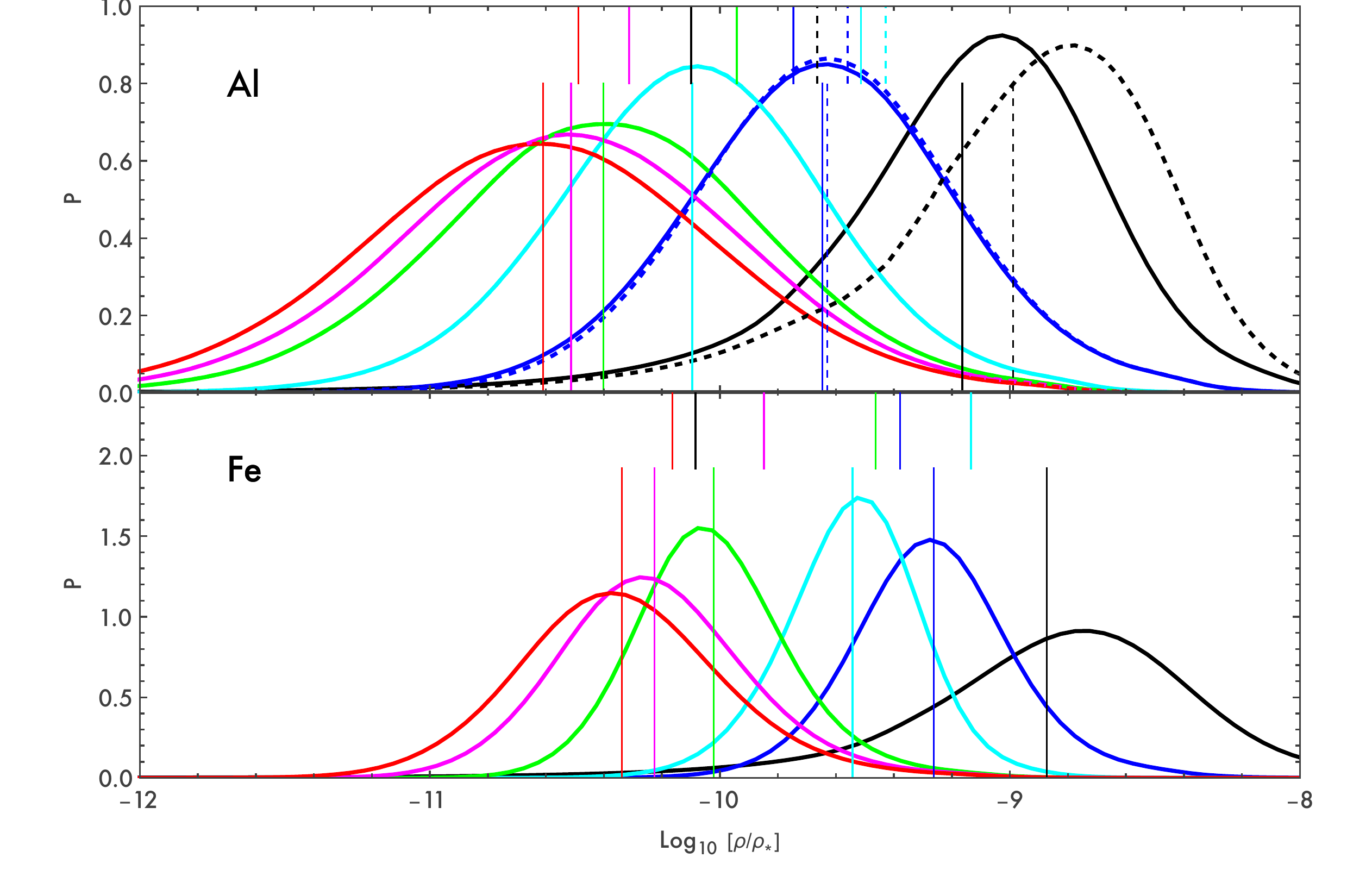}
\end{center}
\caption{Instantaneous star formation model probability functions for
  the logarithmic values of SLR densities at stellar sites for Al (top
  panel - dashed curves include wind) and Fe (lower panel) at 5
  (black), 10 (blue), 15 (light blue), 20 (green), 25 (magenta) and 30
  (red) Myrs.  The lower vertical lines denote the mean value of the
  corresponding distribution.  The upper solid vertical lines denote
  the mean value of the corresponding distribution for our baseline
  model (as shown by the lower vertical lines in Figure 9).  All
  densities are in units of the mean stellar site density $\rho_* =
  150\;M_\odot \;{\rm pc^{-3}}$.  Note that no SN occurred in many
  realizations of our model at 5 Myr, and these results, while not
  shown in the figure, are included in the calculation of the mean
  value for the corresponding distributions.}
\end{figure}

\subsection{Sequential Star Formation}

In this scenario, we consider the possibility that star formation
sweeps across the GMC (e.g., \citealt{elmlada1977}). For the sake of
definiteness, we implement this scenario by correlating the time that
star formation begins within a given cluster to the relative
center-of-mass position of that cluster along the longest axis of the
GMC, and set the sweeping time to 10 Myrs.  All other model parameters
are the same as for our baseline model. Once star formation begins
within a cluster, it proceeds with equal probability over a time
interval $\Delta t = 2$ Myrs, resulting in the same stellar-birth
distribution function for the entire GMC as our baseline model.  As
such, the total SLR mass enrichment as a function of time is identical
to that shown in Figure 4.  However, as shown in Figure 14, the
correlation between stellar position and stellar birth generally leads
to slightly lower values for the probability functions of
$\log[\rho_{Al}/\rho_*]$ and $\log[\rho_{Fe}/\rho_*]$ at star forming
sites (the exception being at 5 Myrs).

\begin{figure}[t]
\begin{center}
\includegraphics[width=0.8\linewidth]{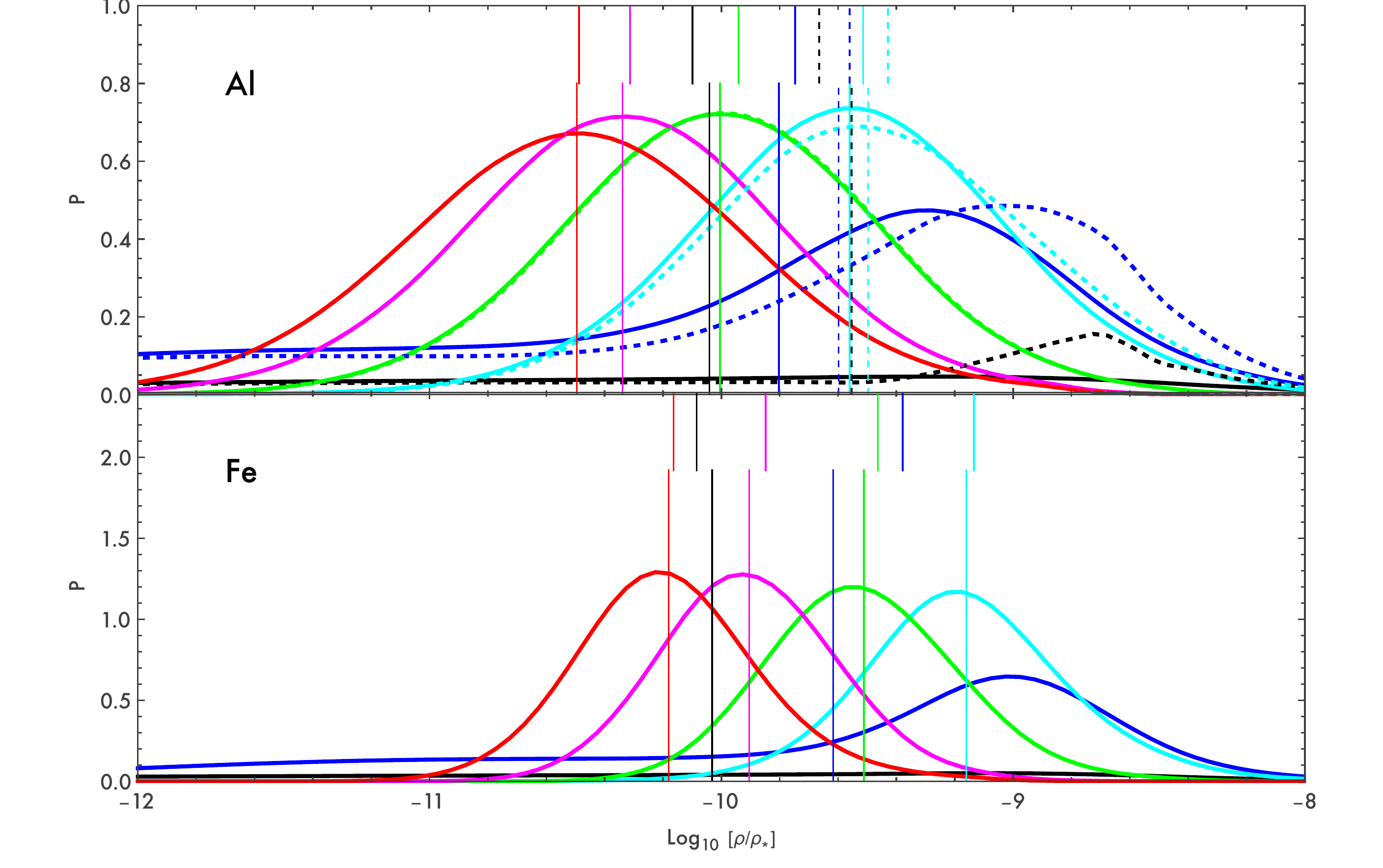}
\end{center}
\caption{Sequential star formation model probability functions for the
  logarithmic values of SLR densities at stellar sites for Al (top
  panel - dashed curves include wind) and Fe (lower panel) at 5
  (black), 10 (blue), 15 (light blue), 20 (green), 25 (magenta) and 30
  (red) Myrs.  The lower vertical lines denote the mean value of the
  corresponding distribution.  The upper solid vertical lines denote
  the mean value of the corresponding distribution for our baseline
  model (as shown by the lower vertical lines in Figure 9).  All
  densities are in units of the mean stellar site density $\rho_* =
  150\;M_\odot \;{\rm pc^{-3}}$.  Note that no SN occurred in many
  realizations of our model at 5 Myr, and these results, while not
  shown in the figure, are included in the calculation of the mean
  value for the corresponding distributions.}
\end{figure}

\subsection{Cloud Structure Variations}

In the standard model, stars are placed within clusters, which in turn
are placed within the GMC using a scheme that creates a structure with
fractal dimension $d = 2$.  We now consider what effect cloud
structure has on our results by considering opposite extremes.
Specifically, we first consider a model where the same scheme is used
to place clusters within the GMC and stars within clusters as for the
baseline model, but with a fractal dimension of $d = 1.6$ for placing
both the clusters in the cloud and the stars within the cluster.  This
scenario leads to an increase in the sub-structure within the GMC.  At
the other extreme, we consider a model where all stars are placed
randomly within the GMC in absence of any clusters.

For the lower fractal dimension scenario, star formation evolves in
the same manner as our baseline model. Specifically, star formation
within a given cluster begins at a time chosen randomly between $t = 0
- 10$ Myr, and proceeds with equal probability over a time interval of
$\Delta t =2\,{\rm Myrs}$.  For consistency, stellar birth in the
random placement model is assigned through the same stellar birth
distribution function that is used to describe the scenario with lower
fractal dimension (which in turn is the same as for our baseline
model). All other model parameters are the same as for our baseline
model.

As illustrated in Figure 15, lowering the fractal dimension from $d =
2$ to $d = 1.6$ has a minimal effect on how SLRs are distributed
throughout the GMC. The resulting probability distributions are nearly
the same. In contrast, the scenario with randomly placed stars (shown
in Figure 16) leads to lower SLR abundances at stellar sites. In this
random model, the stellar locations are spread out more evenly over
the entire GMC and the supernovae themselves are less clustered. As a
result, the random distribution of stellar locations leads to less
concentration of the SLRs and a corresponding deficit of high
abundances.

\begin{figure}[t]
\begin{center}
\includegraphics[width=0.8\linewidth]{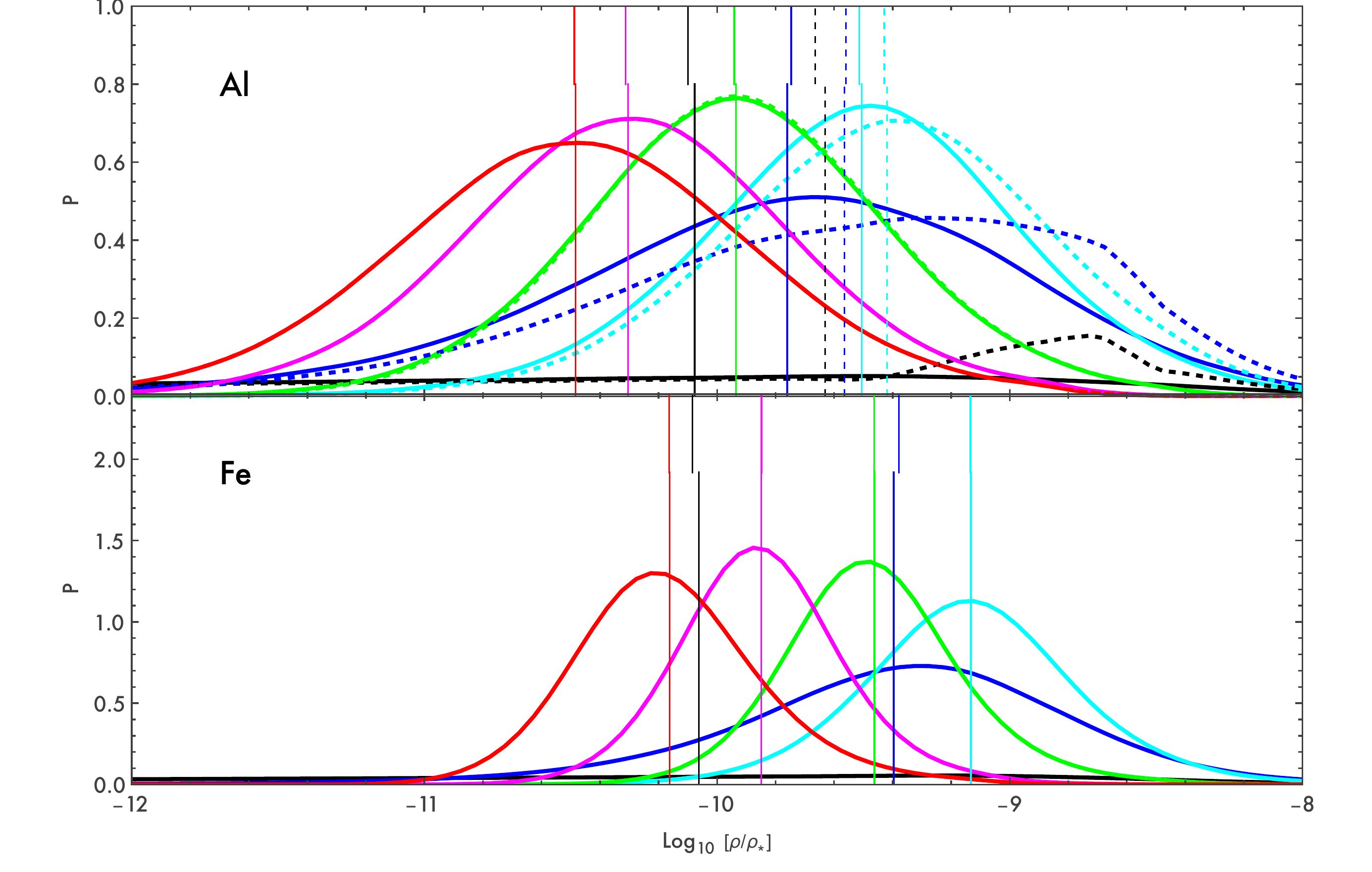}
\end{center}
\caption{High fractal structure ($d = 1.6$) model probability
  functions for the logarithmic values of SLR densities at stellar
  sites for Al (top panel - dashed curves include wind) and Fe (lower
  panel) at 5 (black), 10 (blue), 15 (light blue), 20 (green), 25
  (magenta) and 30 (red) Myrs.  The lower vertical lines denote the
  mean value of the corresponding distribution.  The upper solid
  vertical lines denote the mean value of the corresponding
  distribution for our baseline model (as shown by the lower vertical
  lines in Figure 9).  All densities are in units of the mean stellar
  site density $\rho_* = 150\;M_\odot \;{\rm pc^{-3}}$.  Note that no
  SN occurred in many realizations of our model at 5 Myr, and these
  results, while not shown in the figure, are included in the
  calculation of the mean value for the corresponding distributions.}
\end{figure}

\begin{figure}[t]
\begin{center}
\includegraphics[width=0.8\linewidth]{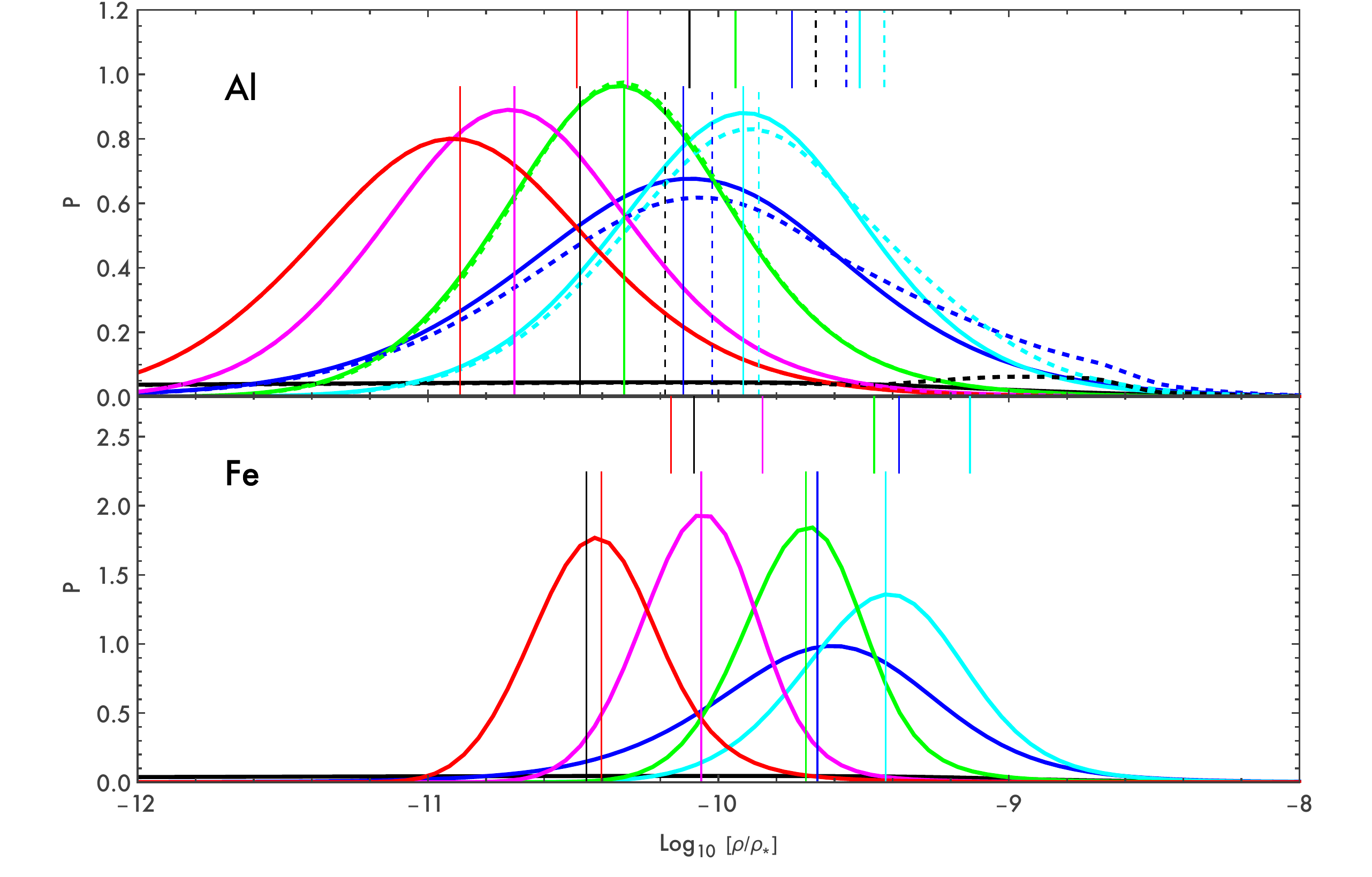}
\end{center}
\caption{Random star location model probability functions for the
  logarithmic values of SLR densities at stellar sites for Al (top
  panel - dashed curves include wind) and Fe (lower panel) at 5
  (black), 10 (blue), 15 (light blue), 20 (green), 25 (magenta) and 30
  (red) Myrs.  The lower vertical lines denote the mean value of the
  corresponding distribution.  The upper solid vertical lines denote
  the mean value of the corresponding distribution for our baseline
  model (as shown by the lower vertical lines in Figure 9).  All
  densities are in units of the mean stellar site density $\rho_* =
  150\;M_\odot \;{\rm pc^{-3}}$.  Note that no SN occurred in many
  realizations of our model at 5 Myr, and these results, while not
  shown in the figure, are included in the calculation of the mean
  value for the corresponding distributions.}
\end{figure}

\subsection{Smaller Cluster Membership} 

While the cluster distribution function chosen for our baseline model
is based on our current understanding of star formation, local
clusters (within $1 - 2$ kpc) have been observed to have memberships
that range from $N$ = 30 to 2000 \citep{ladalada,porras}.  Motivated
by this local sample, we also consider the scenario in which the
cluster membership $N$ spans a smaller range from $10^2$ to $10^3$.
All other model parameters are the same as for our baseline model.

Within the smaller range of stellar membership size $N$, clusters are
drawn from the (usual) distribution function given by equation
(\ref{clusterdist}). As before, we sample the distribution until the
collective stellar content reaches $N^*_{mc}$, with the last drawn
cluster's membership reduced as necessary to achieve that outcome.
For this set of assumptions, the mean cluster membership size is only
$\langle N\rangle = 256$, so that a giant molecular cloud is expected
to have $\sim 200$ clusters with radii $R_c = 0.7 - 1.5$ pc.  We note
that for our assumed cluster distribution, half of the GMC stars are
expected to belong to clusters with membership $N\le 316$.

As shown in Figure 17, the resulting distributions are shifted toward
lower values compared to the standard case. The smaller clusters,
which are more populous, act to distribute the stars more evenly
across the cloud (but not as evenly as in the case of the random
distribution model of Section 3.3). As a result, fewer locations are
near multiple supernovae, and the supernovae themselves are more
spread out, so that fewer locations have high abundances of SLRs.

\begin{figure}[t]
\begin{center}
\includegraphics[width=0.8\linewidth]{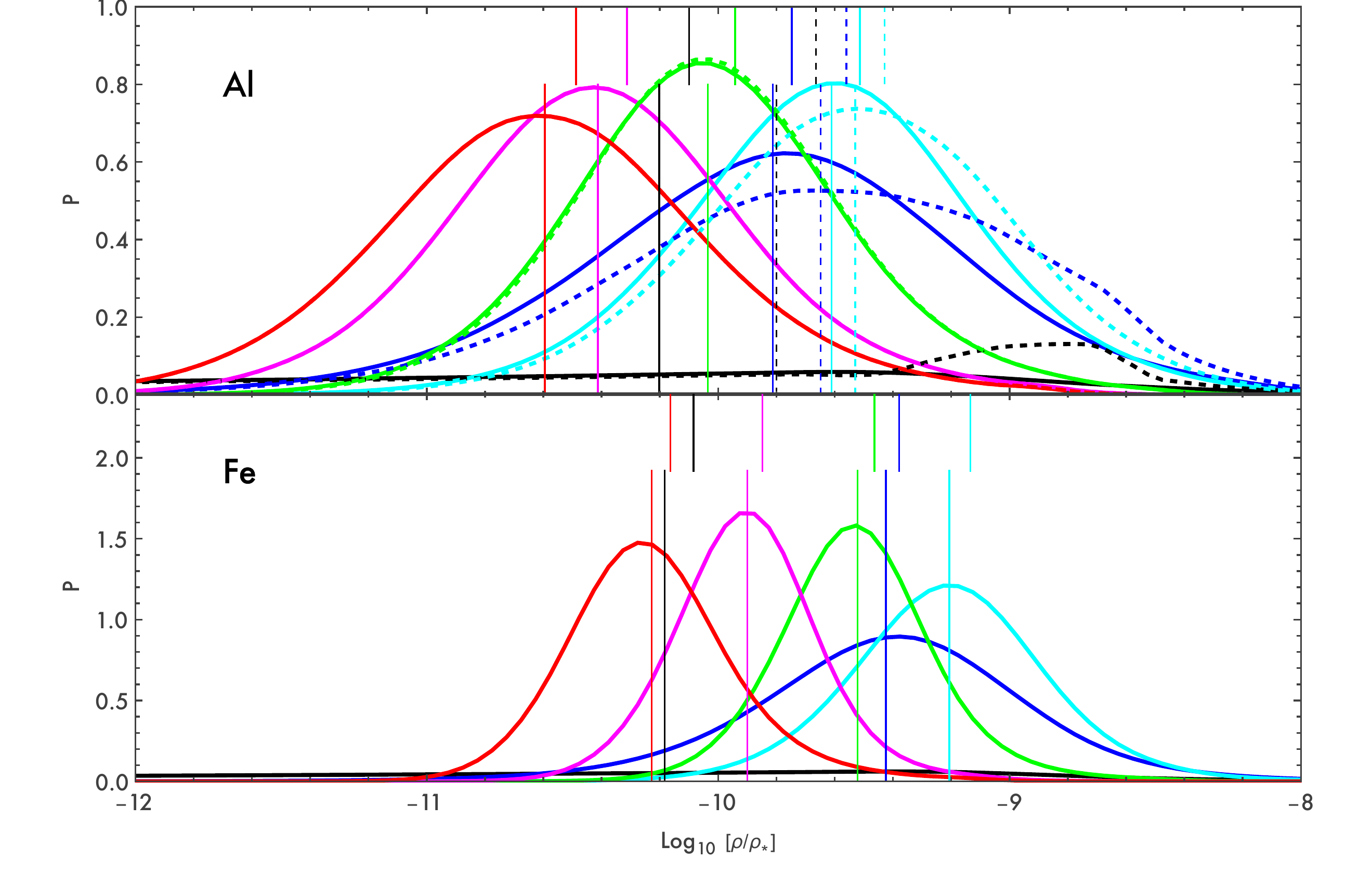}
\end{center}
\caption{Small cluster model probability functions for the logarithmic
  values of SLR densities at stellar sites for Al (top panel - dashed
  curves include wind) and Fe (lower panel) at 5 (black), 10 (blue),
  15 (light blue), 20 (green), 25 (magenta) and 30 (red) Myrs.  The
  lower vertical lines denote the mean value of the corresponding
  distribution.  The upper solid vertical lines denote the mean value
  of the corresponding distribution for our baseline model (as shown
  by the lower vertical lines in Figure 9).  All densities are in
  units of the mean stellar site density $\rho_* = 150\;M_\odot \;{\rm
    pc^{-3}}$.  Note that no SN occurred in many realizations of our
  model at 5 Myr, and these results, while not shown in the figure,
  are included in the calculation of the mean value for the
  corresponding distributions.}
\end{figure}

\subsection{Propagation Parameters}

As noted in \S 2.3, particle transport in GMC environments is poorly
constrained both observationally and theoretically.  We can address
this issue by considering varying values for the diffusion constant,
which determines the manner in which the SLRs are transported across
the molecular cloud. Given the time scales of interest ($t\sim10$ Myr)
and the standard value of the diffusion constant ($D$ = 10 (km/s)pc),
the typical transport distances are comparable to the typical sizes of
molecular clouds ($r\sim\sqrt{Dt}\sim10$ pc).  As a result, larger
values of the diffusion constant allow for a substantial fraction of
the SLRs to be carried out of the GMC and hence become unavailable for
enriching forming stars at later epochs. In contrast, smaller values
lead to much less transport and allow for enhanced localized enrichment.

Here we first consider the case where the diffusion constant has the
smaller value $D$ = 1 (km/s)pc, as shown in Figure 18.  The resulting
distributions of SLRs show enhanced abundances due to the smaller
diffusion constant. In this scenario, the SLRs stay relatively close
to their birth clusters and are effective at enriching stars forming
in the vicinity. This enhancement effect is larger for $^{60}$Fe
compared to $^{26}$Al due to its longer half-life.

For larger values of the diffusion constant, $D=100$ (km/s)pc, Figure
19 shows that the SLR abundances are significantly lower than in the
standard baseline model. This result is expected, as the larger
diffusion constant allows for a substantial fraction of the SLRs to be
swept out of the cloud.

Note that of all the parameters varied in this section, the value of
the diffusion coefficient is the least constrained. With its possible
range of two orders of magnitude, this quantity has the largest effect
on the resulting distributions of SLRs abundances.

\begin{figure}[t]
\begin{center}
\includegraphics[width=0.8\linewidth]{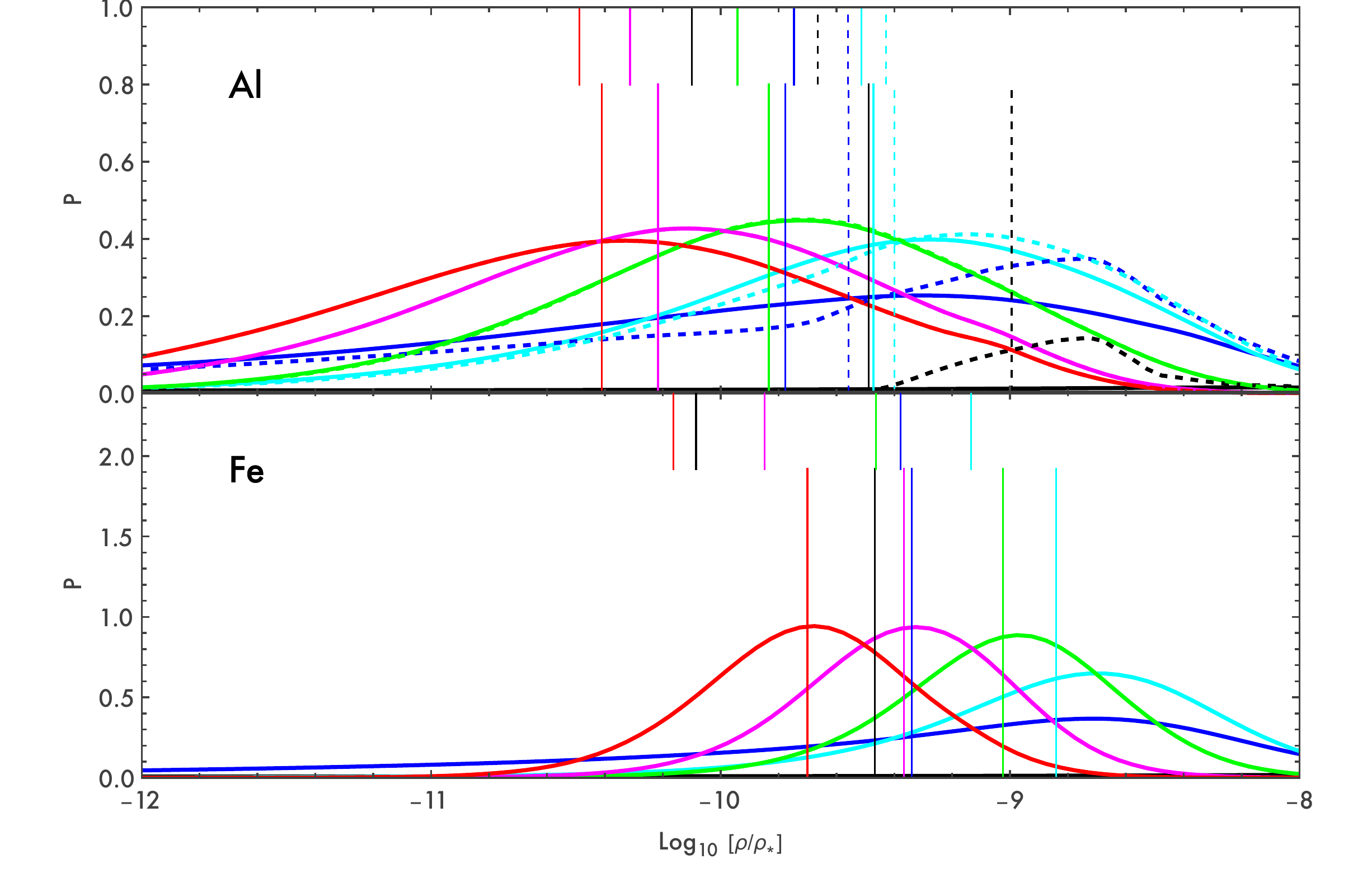}
\end{center}
\caption{$D=1\,({\rm km/s)pc}$ model probability functions for the
  logarithmic values of SLR densities at stellar sites for Al (top
  panel - dashed curves include wind) and Fe (lower panel) at 5
  (black), 10 (blue), 15 (light blue), 20 (green), 25 (magenta) and 30
  (red) Myrs.  The lower vertical lines denote the mean value of the
  corresponding distribution.  The upper solid vertical lines denote
  the mean value of the corresponding distribution for our baseline
  model (as shown by the lower vertical lines in Figure 9).  All
  densities are in units of the mean stellar site density $\rho_* =
  150\;M_\odot \;{\rm pc^{-3}}$.  Note that no SN occurred in many
  realizations of our model at 5 Myr, and these results, while not
  shown in the figure, are included in the calculation of the mean
  value for the corresponding distributions.}
\end{figure}

\begin{figure}[t]
\begin{center}
\includegraphics[width=0.8\linewidth]{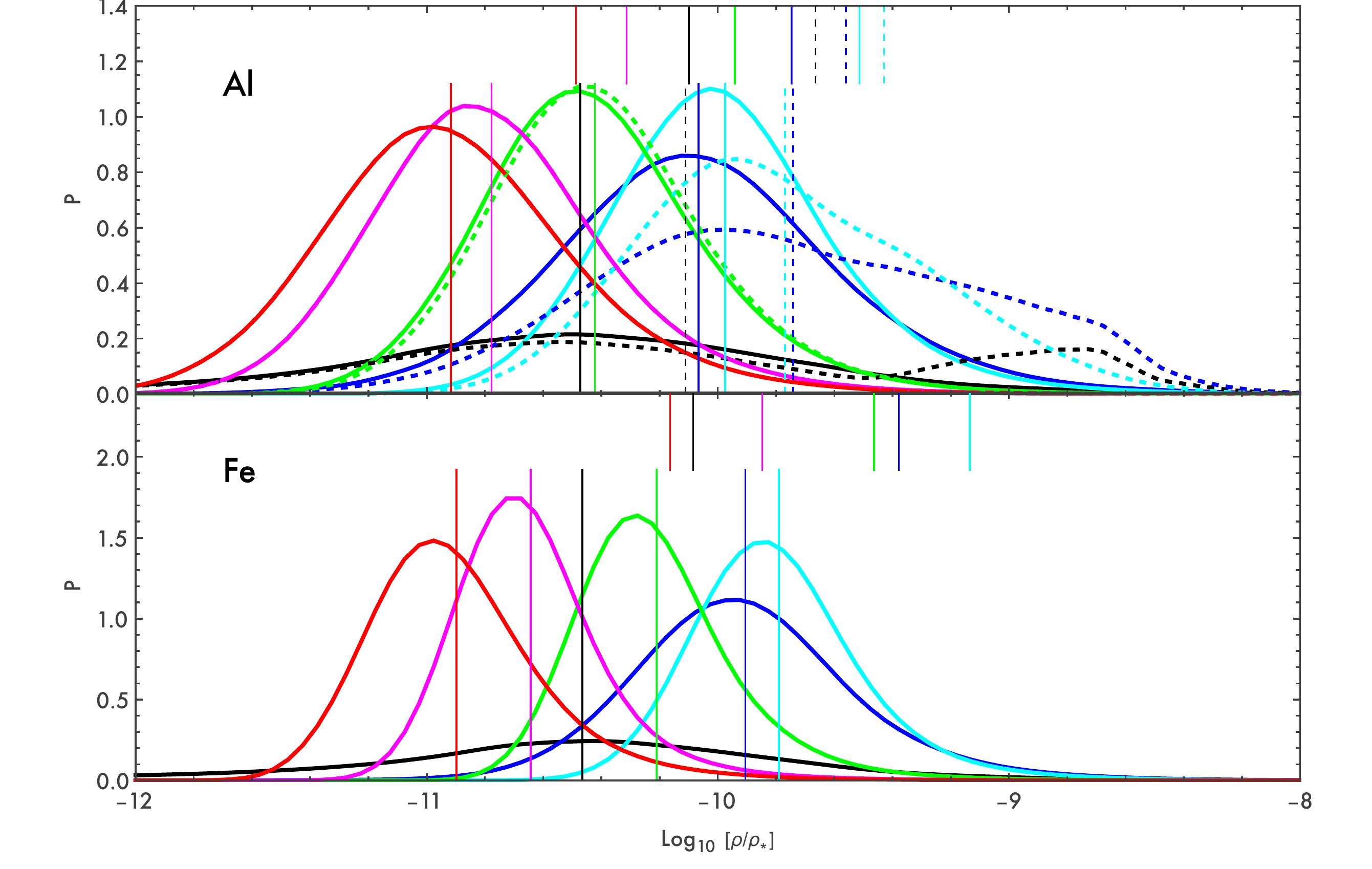}
\end{center}
\caption{$D=100\,({\rm km/s)pc}$ model probability functions for the
  logarithmic values of SLR densities at stellar sites for Al (top
  panel - dashed curves include wind) and Fe (lower panel) at 5
  (black), 10 (blue), 15 (light blue), 20 (green), 25 (magenta) and 30
  (red) Myrs.  The lower vertical lines denote the mean value of the
  corresponding distribution.  The upper solid vertical lines denote
  the mean value of the corresponding distribution for our baseline
  model (as shown by the lower vertical lines in Figure 9).  All
  densities are in units of the mean stellar site density $\rho_* =
  150\;M_\odot \;{\rm pc^{-3}}$.  Note that no SN occurred in many
  realizations of our model at 5 Myr, and these results, while not
  shown in the figure, are included in the calculation of the mean
  value for the corresponding distributions.}
\end{figure}

\subsection{Anisotropic SLR Transport}  

The use of a single diffusion coefficient to describe particle
transport in a GCM is highly idealized.  Indeed, recent insights about
the evolution of molecular clouds indicate that massive-star feedback
erodes clouds within a few Myrs, and that wind-blown bubbles are
important sites where radioisotopes spread.  Far from homogeneous,
GMCs are highly nonuniform, and exhibit thin filamentary structures
extending over $\sim$100 pc. The spherical injection and propagation
models adopted in our work are therefore too simplistic to fully
describe the transport of matter in such a hot and stirred
massive-star environment.
 
While creating a particle-transport model that takes into account the
structural complexities of GMCs is beyond the scope of our work, we
model the effects on non-uniformity in the GMC environment by
selecting the diffusion coefficient used to determine the density at a
given stellar site (via equation [19]) from a distribution function
$F(D)$.  In order to generate this distribution function, we first
calculate the column density along a 4 pc line of sight extending in
six directions from the position of each field star for one
realization of our GMC.  Doing so yields $\approx 300,000$ values of
column densities whose distribution characterizes the non-uniform
structure of our GMC environments.  Since the diffusion constant is
expected to be inversely proportional to the mean free path, we then
generate a corresponding distribution of diffusion constants through
the relation $D_i= 10 \langle N \rangle /N_i$ pc, where $\langle N
\rangle$ is the mean of the column density distribution, and $N_i$ is
the $ith$ member of the column density distribution.  The resulting 
distribution of diffusion constants $F(D)$ is shown in Figure 20.
 
The spread in $F(D)$ indicates that particle diffusion will not
proceed uniformly in all directions, but rather will vary in
accordance with the diffusion constant distribution. Note that the
width of the distribution is approximately a factor of 2 on either
side of the mean. With this result, we perform a calculation using our
baseline model parameters, but with the diffusion constant $D$ used to
calculate the SLR densities at a each star-formation site (via
eq. [19]) randomly selected form our generated distribution function
$F(D)$.  The results of this procedure are shown in Figure 21 (solid
curves) along with the corresponding distributions from the baseline
model (dotted curves) for which $D = 10$ km/s pc (as presented in
Figure 9).  The inclusion of non-uniform propagations slightly
broadens the density distribution functions (as expected) with
slightly lower mean values. However, as can be seen by comparing the
solid and dotted curves, the overall effect for our model is
quite modest.  We also note that filamentary clouds allow for
more complicated transport behavior, wherein gas streaming away from
one filament can reach other filaments. As a result, future work
should explore this issue further, including a dynamical cloud medium
shaped by feedback of winds and explosions. 
  
\begin{figure}[t]
\begin{center}
\includegraphics[width=0.8\linewidth]{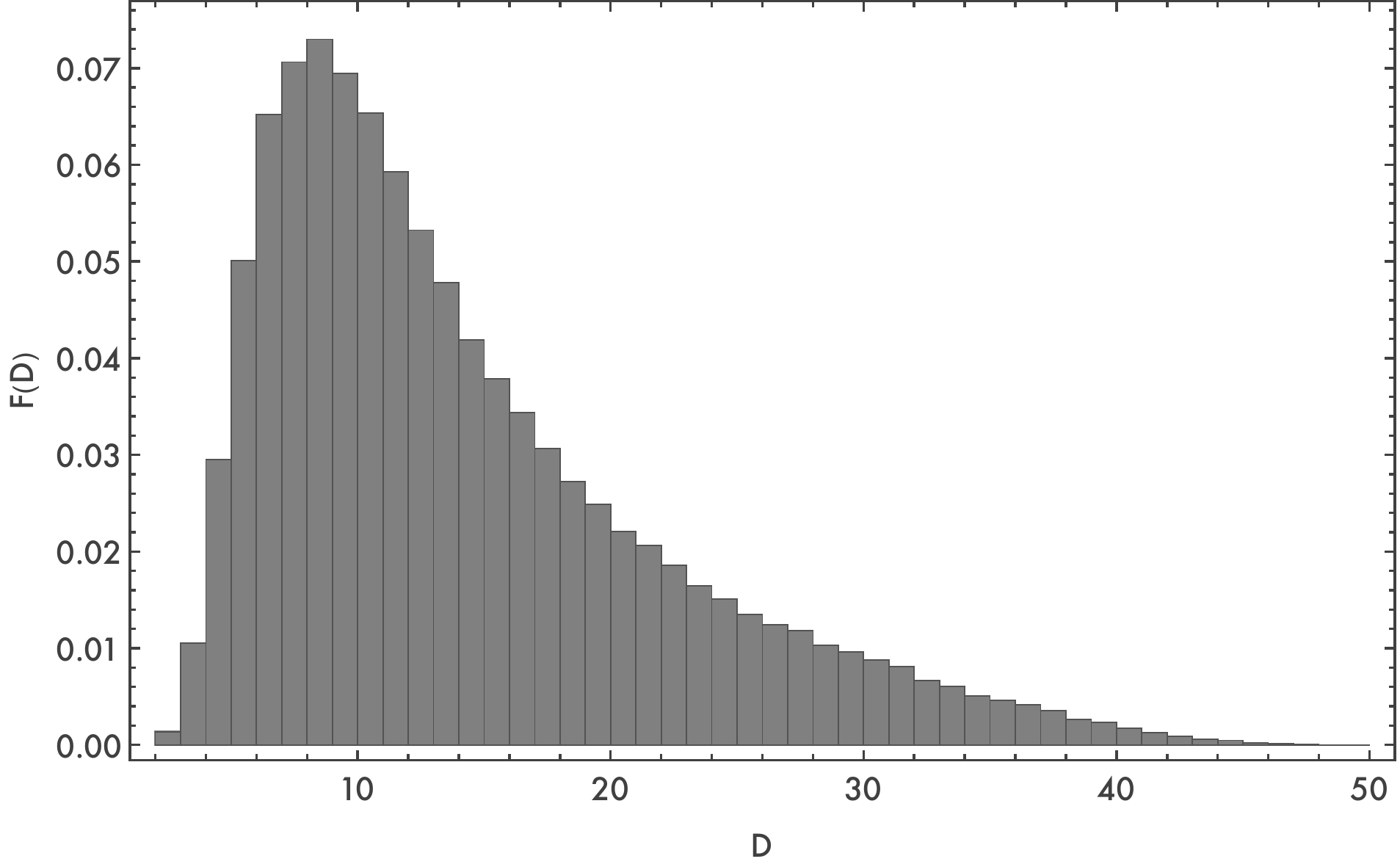}
\end{center}
\caption{The distribution function of diffusion constants built
  through the relation $D_i= 10 \langle N \rangle /N_i$ pc, where
  $N_i$ is the value of the $ith$ member of the distribution of column
  densities calculated along a 4 pc line of sight extending in six
  directions from the position of each field star for one realization
  of our GMC, and $\langle N \rangle$ is the mean of the column
  density distribution. }
\end{figure}

\begin{figure}[t]
\begin{center}
\includegraphics[width=0.8\linewidth]{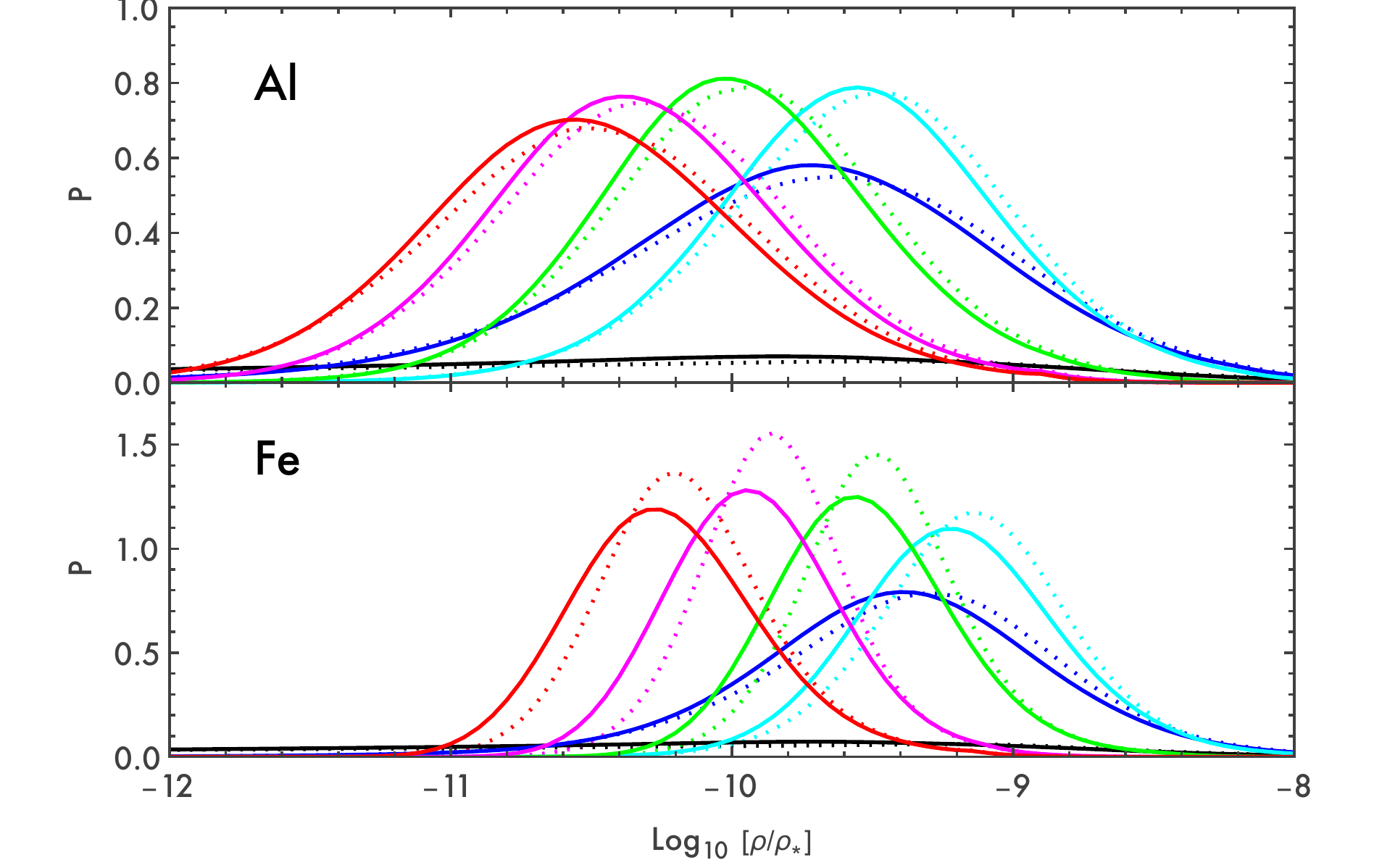}
\end{center}
\caption{Non-uniform model probability functions (solid curves) for
  the logarithmic values of SLR densities at stellar sites for Al (top
  panel) and Fe (lower panel) at 5 (black), 10 (blue), 15 (light
  blue), 20 (green), 25 (magenta) and 30 (red) Myrs.  Dotted curves
  show the corresponding distribution for our baseline model (as shown
  in Figure 9).  All densities are in units of the mean stellar site
  density $\rho_* = 150\;M_\odot \;{\rm pc^{-3}}$.  Note that no SN
  occurred in many realizations of our model at 5 Myr, and these
  results, while not shown in the figure, are included in the
  calculation of the mean value for the corresponding distributions.}
\end{figure}

\medskip 
\section{Conclusions} 
\label{sec:conclude} 

Given the possible importance of short-lived radionuclides during the
processes of star and planet formation, this paper considers the
expected distributions of SLRs on the size and mass scales of
molecular clouds.  Supernovae (for both $^{60}$Fe and $^{26}$Al) and
stellar winds (for $^{26}$Al only) provide SLRs for the star forming
population produced within the cloud. The main result of this work is
the distribution of possible radioactive enrichment levels for forming
stars and planets. The resulting values for the mass fractions of
$^{26}$Al are listed in Table 2 for a collection of cloud ages and for
the different scenarios considered in this work. Table 3 shows the
corresponding mass fractions for $^{60}$Fe.

\subsection{Summary of Results} 
\label{sec:summary} 

Our main conclusions can be summarized as follows:

\medskip 

[1] Distributed enrichment of SLRs from supernova sources produce a
wide distribution of abundances of radioactive nuclei across molecular
clouds. For the isotope $^{60}$Fe, we find abundance levels in the
approximate range $\rho_{SLR}/\rho_\ast\sim10^{-11}-10^{-8}$, where
such enrichment is expected during the first $\sim10$ Myr of star
formation. The corresponding enrichment levels for $^{26}$Al are
somewhat smaller but roughly comparable (see Figure 9). The
distributions of SLR abundances are time dependent, and decrease
steadily after the epoch of star formation has finished (here, for
times later than 12 Myr).

\medskip 

[2] The ratio of $^{60}$Fe/$^{26}$Al also displays a wide distribution
(Figure 10), with typical values somewhat greater than unity
($\sim1-2$).  Significantly, supernovae lead to greater enrichment
levels for $^{60}$Fe, compared to $^{26}$Al, which is opposite to the
trend observed/inferred for our solar system (and for the galaxy as a
whole). This trend arises because the two isotopes are produced with
roughly comparable abundances in supernovae, whereas $^{60}$Fe has a
significantly longer half-life (2.6 Myr, compared to 0.72 Myr for
$^{26}$Al). These results correspond to the typical values sampled
from the entire distributions found by this paper. Any particular
realization still could have more $^{26}$Al than $^{60}$Fe. For
example, the scenario with efficient SLR propagration (large diffusion
constant) and evaluated at late evolution times allows for larger
values of $^{26}$Al/$^{60}$Fe (see Figure 19).

\medskip 

[3] The distributions for the absolute enrichment levels indicate that
this GMC-wide enrichment scenario could explain the absolute values of
the abundances of SLRs inferred for the early Solar System. The
probability of attaining the observed Solar System mass fractions
depends on time (as determined by the star formation history of the
cloud). For our standard scenario, forming stars have a $\sim10\%$
chance of reaching $^{26}$Al enrichment levels for cloud times $t$ =
10 -- 15 Myr, with considerably lower probabilities outside that span
of time.  On the other hand, the distributions of abundance ratios 
$^{60}$Fe/$^{26}$Al are generally not consistent with current estimates.

\medskip 

[4] In addition to our baseline enrichment scenario (Figure 9), we
have varied the input parameters of the model.  Specifically, this
paper provides SLR distributions for instantaneous star formation
(Figure 13), sequential star formation (Figure 14), different fractal
dimension for cloud structure (Figure 15), uniform-random distribution
of star formation sties (Figure 16), smaller clusters (Figure 17),
varying SLR propagation efficiencies (Figures 18 and 19), and for the
effects of non-spherical propagation (Figures 20 and 21). Over most of
this parameter space, the distributions of SLRs display only modest
variations. The most important variable --- in terms of changing the
distributions of SLRs abundances --- is the diffusion coefficient for
the propagation of SLRs through the molecular cloud. If the diffusion
constant becomes sufficiently large ($D\gta30$ pc$^2$ Myr$^{-1}$),
then the enrichment levels decrease substantially. The key issue
required for maintaining significant levels of nuclear enrichment is
the retention of the SLRs within the molecular cloud.

\medskip 

\begin{table}
\begin{center}
\begin{tabular}{cccccccc}
\tableline\tableline
Model& Fig. & 5\,{\rm Myrs}& 10\,{\rm Myrs} & 15\,{\rm Myrs} & 20\,{\rm Myrs}& 25\,{\rm Myrs} & 30\,{\rm Myrs} \\
\tableline
Standard &9& $8.0\times 10^{-11}$ &$1.8\times 10^{-10}$ &$3.1\times 10^{-10}$ &$1.1\times 10^{-10}$ &$4.9\times 10^{-11}$  &$3.3\times 10^{-11}$  \\
Inst. SF &13& $6.8\times 10^{-10}$ &$2.3\times 10^{-10}$ &$8.0\times 10^{-11}$ &$4.0\times 10^{-11}$ &$3.1\times 10^{-11}$  &$2.5\times 10^{-11}$  \\
Seq. SF &14& $9.1\times 10^{-11}$ &$1.6\times 10^{-10}$ &$2.8\times 10^{-10}$ &$9.9\times 10^{-11}$ &$4.6\times 10^{-11}$  &$3.2\times 10^{-11}$  \\
$d=1.6$ &15& $8.4\times 10^{-11}$ &$1.7\times 10^{-10}$ &$3.1\times 10^{-10}$ &$1.2\times 10^{-10}$ &$5.0\times 10^{-11}$  &$3.3\times 10^{-11}$  \\
Random &16& $3.3\times 10^{-11}$ &$7.6\times 10^{-11}$ &$1.2\times 10^{-10}$ &$4.7\times 10^{-11}$ &$2.0\times 10^{-11}$  &$1.3\times 10^{-11}$  \\
Small Cluster&17& $6.3\times 10^{-11}$ &$1.5\times 10^{-10}$ &$2.4\times 10^{-10}$ &$9.2\times 10^{-11}$ &$3.9\times 10^{-11}$  &$2.5\times 10^{-11}$  \\
$D = 1$ km/s/pc &18& $3.2\times 10^{-10}$ &$1.7\times 10^{-10}$ &$3.4\times 10^{-10}$ &$1.5\times 10^{-10}$ &$6.1\times 10^{-11}$  &$3.9\times 10^{-11}$  \\
$D = 10^2$ km/s/pc &19& $3.4\times 10^{-11}$ &$8.6\times 10^{-11}$ &$1.1\times 10^{-10}$ &$3.8\times 10^{-11}$ &$1.7\times 10^{-11}$  &$1.2\times 10^{-11}$  \\
Nonuniform&21& $7.0\times 10^{-11}$ &$1.7\times 10^{-10}$ &$2.8\times 10^{-10}$ &$1.0\times 10^{-10}$ &$4.5\times 10^{-11}$  &$3.0\times 10^{-11}$  \\
 \tableline
\end{tabular}
\caption{Mean values of the density of $^{26}$Al produced by SN events at the star formation sites at 6 different GMC ages for the different models
explored in this work.  All densities are in units of the mean stellar site density $\rho_* = 150\;M_\odot \;{\rm pc^{-3}}$.  Wind produced  $^{26}$Al is excluded.}
\end{center}
\end{table}

\begin{table}
\begin{center}
\begin{tabular}{cccccccc}
\tableline\tableline
Model& Fig. & 5\,{\rm Myrs}& 10\,{\rm Myrs} & 15\,{\rm Myrs} & 20\,{\rm Myrs}& 25\,{\rm Myrs} & 30\,{\rm Myrs} \\
\tableline
Standard &9& $8.2\times 10^{-11}$ &$4.2\times 10^{-10}$ &$7.3\times 10^{-10}$ &$3.4\times 10^{-10}$ &$1.4\times 10^{-10}$  &$6.9\times 10^{-11}$  \\
Inst. SF &13& $1.3\times 10^{-9}$ &$5.5\times 10^{-10}$ &$2.9\times 10^{-10}$ &$9.5\times 10^{-11}$ &$6.0\times 10^{-11}$  &$4.6\times 10^{-11}$  \\
Seq. SF &14& $9.3\times 10^{-11}$ &$2.4\times 10^{-10}$ &$6.9\times 10^{-10}$ &$3.1\times 10^{-10}$ &$1.2\times 10^{-10}$  &$6.6\times 10^{-11}$  \\
$d=1.6$ &15& $8.7\times 10^{-11}$ &$4.0\times 10^{-10}$ &$7.4\times 10^{-10}$ &$3.4\times 10^{-10}$ &$1.4\times 10^{-10}$  &$6.9\times 10^{-11}$  \\
Random &16& $3.5\times 10^{-11}$ &$2.2\times 10^{-10}$ &$3.8\times 10^{-10}$ &$2.0\times 10^{-10}$ &$8.7\times 10^{-11}$  &$4.0\times 10^{-11}$  \\
Small Cluster&17& $6.5\times 10^{-11}$ &$3.8\times 10^{-10}$ &$6.2\times 10^{-10}$ &$3.0\times 10^{-10}$ &$1.3\times 10^{-10}$  &$5.9\times 10^{-11}$  \\
$D = 1$ km/s/pc &18& $3.4\times 10^{-10}$ &$4.6\times 10^{-10}$ &$1.4\times 10^{-9}$ &$9.4\times 10^{-10}$ &$4.3\times 10^{-10}$  &$2.0\times 10^{-10}$  \\
$D = 10^2$ km/s/pc &19& $3.4\times 10^{-11}$ &$1.2\times 10^{-10}$ &$1.6\times 10^{-10}$ &$6.2\times 10^{-11}$ &$2.3\times 10^{-11}$  &$1.3\times 10^{-11}$  \\
Nonuniform&21& $7.3\times 10^{-11}$ &$3.8\times 10^{-10}$ &$6.4\times 10^{-10}$ &$3.0\times 10^{-10}$ &$1.2\times 10^{-10}$  &$5.9\times 10^{-11}$  \\

 \tableline
\end{tabular}
\caption{Mean values of the density of $^{60}$Fe produced by SN events at the star formation sites at 6 different GMC ages for the different models
explored in this work.  All densities are in units of the mean stellar site density $\rho_* = 150\;M_\odot \;{\rm pc^{-3}}$.}
\end{center}
\end{table}

\subsection{Discussion} 
\label{sec:discuss} 

In terms of the absolute SLR abundances that affect forming stars and
planets, we find intermediate results. The expected nuclear enrichment
is large enough to provide significant sources of heating and
ionization for circumstellar disks and planets, but is not large
enough to completely dominate the picture (see the discussion of
\citealt{lichtenberg19,reiter2020}, and references therein).
Moreover, the distributions of SLRs abundances are wide, roughly
comparable to their mean values. As a result, enrichment levels
must be described in terms of probability distributions, i.e., we cannot 
assign a single, typical value to the expected degree of radioactive 
enrichment. 

Although the focus of this paper is to provide the distributions of
SLRs for the entire population of stars forming within a cloud, it is
useful to put our Solar System in context. Here, our results present a
somewhat complicated picture for nuclear enrichment scenarios. On one
hand, the probability of reaching the abundance levels for $^{60}$Fe
and $^{26}$Al inferred for our Solar System are high enough that the
observed values are not problematic.  In this sense, our Solar System
is not out of the ordinary (consistent with previous work, from
\citealt{jura2013} to \citealt{young2020}). On the other hand, the
abundance ratio of iron to aluminum $^{60}$Fe/$^{26}$Al is generally
larger than unity, rather than smaller than unity as observed for the
Solar System (where $^{60}$Fe/$^{26}$Al $\lta0.25$).  This ratio is
also smaller than unity on galactic scales, $^{60}$Fe/$^{26}$Al
$\sim0.4-0.9$, as measured by gamma-ray lines \citep{wang2020}. 

This study considers the enrichment provided by the transport of SLRs
across molecular clouds, where the sources include supernovae and
stellar winds. Additional sources of radioactive material could also
play a role, including spallation in the interstellar medium
\citep{desch2010} and spallation from local sources of cosmic rays
\citep{shu1997}. Significantly, spallation can provide an additional
source for $^{26}$Al, whereas only stellar nucleosynthesis is known to
produce $^{60}$Fe. As a result, these additional sources of
radioactive nuclei act to decrease the abundance ratio
$^{60}$Fe/$^{26}$Al.

Taken together, the results of this paper suggest that the expected
abundances of both $^{60}$Fe and $^{26}$Al in molecular clouds are
large enough to affect star and planet formation.  Specifically,
enrichment levels comparable to those inferred for the early solar
nebula can be attained with reasonable probability, although such
values fall toward the high end of the distribution. In any case, the
expected enrichment levels provide significant contributions to
ionization and heating in forming stars and planets.  On the other
hand, enrichment through supernovae and stellar winds alone do not
provide a full explanation for all the Solar System SLRs, including
their exact abundance ratios.  In particular, the mass ratios
$^{60}$Fe/$^{26}$Al predicted from this study are generally larger
than estimates for both the early Solar System and the Galaxy as a
whole (although the tail of the distribution includes smaller values).
Additional work is thus necessary to obtain a full understanding of
the SLR abundances for the early Solar System, the Galaxy, and the
current population of forming stars and planets.

\acknowledgments

We thank the referee for an extensive set of constructive comments.
MF is grateful to the Hauck Foundation. FCA is supported in part by
the Leinweber Center for Theoretical Physics at the University of
Michigan.

\end{document}